\documentclass[12pt,a4paper,final]{iopart}

 \expandafter\let\csname equation*\endcsname\relax
  \expandafter\let\csname endequation*\endcsname\relax
  
\usepackage{amsmath}

 \usepackage{iopams}  
\usepackage{graphicx}
\usepackage{braket}

\usepackage[breaklinks=true,colorlinks=true,linkcolor=blue,urlcolor=blue,citecolor=blue]{hyperref}

\begin{document}

\title[Permutation blocking path integral Monte Carlo]{Permutation blocking path integral Monte Carlo: A highly efficient approach to the simulation of strongly degenerate non-ideal fermions}

\author{Tobias Dornheim$^{1}$, Simon Groth$^{1}$, Alexey Filinov$^{1,2}$ and Michael Bonitz$^{1}$}
\address{$^1$Institut f\"ur Theoretische Physik und Astrophysik, Christian-Albrechts-Universit\"at, Leibnizstra{\ss}e 15, Kiel D-24098, Germany}
\address{$^2$Joint Institute for High Temperatures RAS, Izhorskaya Str. 13, 125412 Moscow, Russia}
\ead{dornheim@theo-physik.uni-kiel.de}


\begin{abstract}
Correlated fermions are of high interest in condensed matter (Fermi liquids, Wigner molecules), cold atomic gases and dense plasmas. Here
we propose a novel approach to path integral Monte Carlo (PIMC) simulations of strongly degenerate non-ideal fermions at finite temperature by combining a fourth-order factorization 
of the density matrix with antisymmetric propagators, i.e., determinants, between all imaginary time slices. 
To efficiently run through the modified configuration space, we introduce a modification of the widely used continuous space worm algorithm, which
allows for an efficient sampling at arbitrary system parameters.
We demonstrate how the application of determinants achieves an effective blocking of permutations with opposite signs, leading to a significant relieve of the fermion sign problem.
To benchmark the capability of our method regarding the simulation of degenerate fermions, we consider multiple electrons in a quantum dot and 
compare our results with other ab initio techniques, where they are available.
The present permutation blocking path integral Monte Carlo approach allows us to obtain accurate results even for $N=20$ electrons at low temperature and arbitrary coupling, where no
other ab initio results have been reported, so far.

\end{abstract}



\vspace{2pc}
\submitto{\NJP}


\section{Introduction}
The ab initio simulation of strongly degenerate nonideal fermions at finite temperature is of high current importance for many fields.
The numerous physical applications include electrons in a quantum dot \cite{egger,wigner,rontani,ghosal},
fermionic bilayer systems \cite{fbilayer,ludwig,fbilayer2},
the homogeneous electron gas \cite{brown,prl,vfil4},
dense two-component plasmas \cite{bonitz,morales,proton} in stellar interiors and modern laser compression
experiments (warm dense matter) \cite{fletcher, kraus} and inertial fusion \cite{hurricane}.
Despite remarkable recent progress, existing simulation methods face serious problems.

The widely used path integral Monte Carlo (PIMC) method, e.g.\ \cite{cep}, is a highly successful tool for the ab initio simulation of both distinguishable particles (``boltzmannons``, e.g.\ \cite{mil,clark}) and bosons \cite{cep}
and allows for the calculation of quasi-exact results for up to $N\sim 10^3$ particles \cite{bon} at finite temperature.
However, the application of PIMC to fermions is hampered by the notorious sign problem \cite{loh},
which renders even small systems unfeasible for state of the art techniques and has been revealed to be $NP$-complete for a given representation \cite{troyer}.
With increasing exchange effects, permutation cycles with opposite signs appear with nearly equal frequency and the statistical error increases exponentially.
For this reason, standard PIMC is applicable to fermions only at weak degeneracy, that is, at relatively high temperature or low density.

The recently introduced configuration path integral Monte Carlo (CPIMC) method \cite{tim1, tim2, prl} exhibits a complementary behavior.
This conceptually different approach can be interpreted as a Monte Carlo simulation on a perturbation expansion around the ideal quantum system and, therefore,
CPIMC excells at weak nonideality and strong degeneracy. Unfortunately, the physically most interesting region, where both fermionic exchange and interactions
are strong simultaneously, remains out of reach.

A popular approach to extend standard PIMC to higher degeneracy is Restricted PIMC (RPIMC) \cite{node}, also known as fixed node approximation. 
This idea requires explicit knowledge of the nodal surfaces of the density matrix, which are, in general, unknown and one has to rely on approximations,
thereby introducing an uncontrollable systematic error. In addition, it has been shown analytically \cite{vfil1,vfil2}
that RPIMC does not reproduce the exact density matrix in the limit of the ideal Fermi gas and, therefore, the results become unreliable at increasing degeneracy \cite{prl}.

Recently, DuBois \textit{et al.} \cite{dubois} have suggested that, at least for homogeneous systems, the individual exchange probabilities in PIMC 
are independent of the configuration of other permutations present and that permutation frequencies of large exchange cycles can be extrapolated from few-particle permutations. This would allow for a significant reduction of the configuration space
and a drastic reduction of the sign problem. While first simulation results with this approximation for the short-range interacting $^3$He are in good agreement with experimental data \cite{dubois},
the existing comparison \cite{prl} for long-range Coulomb interaction is insufficient to assess the accuracy and,
in addition, inhomogeneous systems remain out of reach.

Another possibility to relieve the sign problem in fermionic PIMC without introducing any approximations is the usage of antisymmetric imaginary time propagators, i.e., determinants \cite{takahashi, vfil3, vfil4, lyu}.
It is well known that the sign problem becomes more severe with an increasing number of propagators arising from the Trotter-type factorization of the density operator.
Consequently, it has been proposed to combine the antisymmetric propagators with a higher order factorization \cite{suzuki1,suzuki2,sympletic,tia} of the density matrix.
This has recently allowed to obtain an accurate estimate of the ground state energy of degenerate, strongly nonideal electrons in a quantum dot \cite{chin1}.

In the present work, we extend this idea to finite temperature. For this purpose, we combine a fourth-order propagator derived in \cite{chin2}, which has already been succesfully applied to PIMC by Sakkos \textit{et al.}\ \cite{sakkos}, 
with a full antisymmetrization on all time slices to simulate fermions in the canonical ensemble. We demonstrate that the introduction of determinants 
effectively allows for the combination of $N!$ configurations from usual PIMC into a single configuration weight, thereby reducing the complexity of the problem 
and blocking both positive and negative weights to drastically increase the sign.
To efficiently exploit the resulting configuration space with the Metropolis algorithm \cite{metropolis} at arbitrary parameters, we develop 
a set of Monte Carlo updates similar to the usual continuous space worm algorithm (WA) \cite{bon,bon2}.

To demonstrate the capability of our permutation blocking path integral Monte Carlo (PB-PIMC) method, we consider Coulomb interacting fermions in a $2D$ harmonic confinement, cf.\ Eq.\ (\ref{hamilton}),
which can be experimentally realized e.g. by spin-polarized electrons in a quantum dot \cite{egger,wigner,rontani,ghosal}.
\begin{figure}[]
 \centering
 \includegraphics[width=0.49\textwidth]{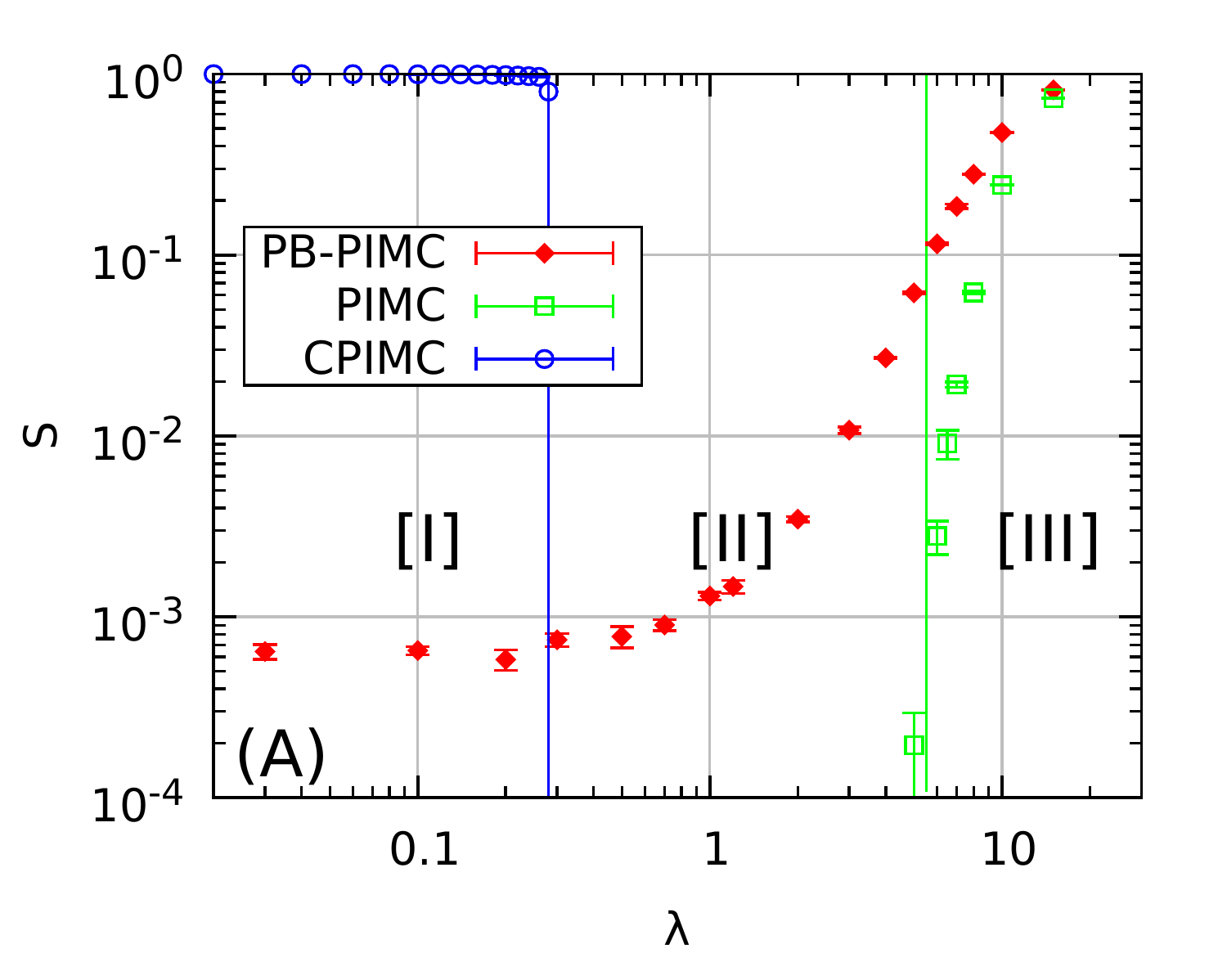}
  \includegraphics[width=0.49\textwidth]{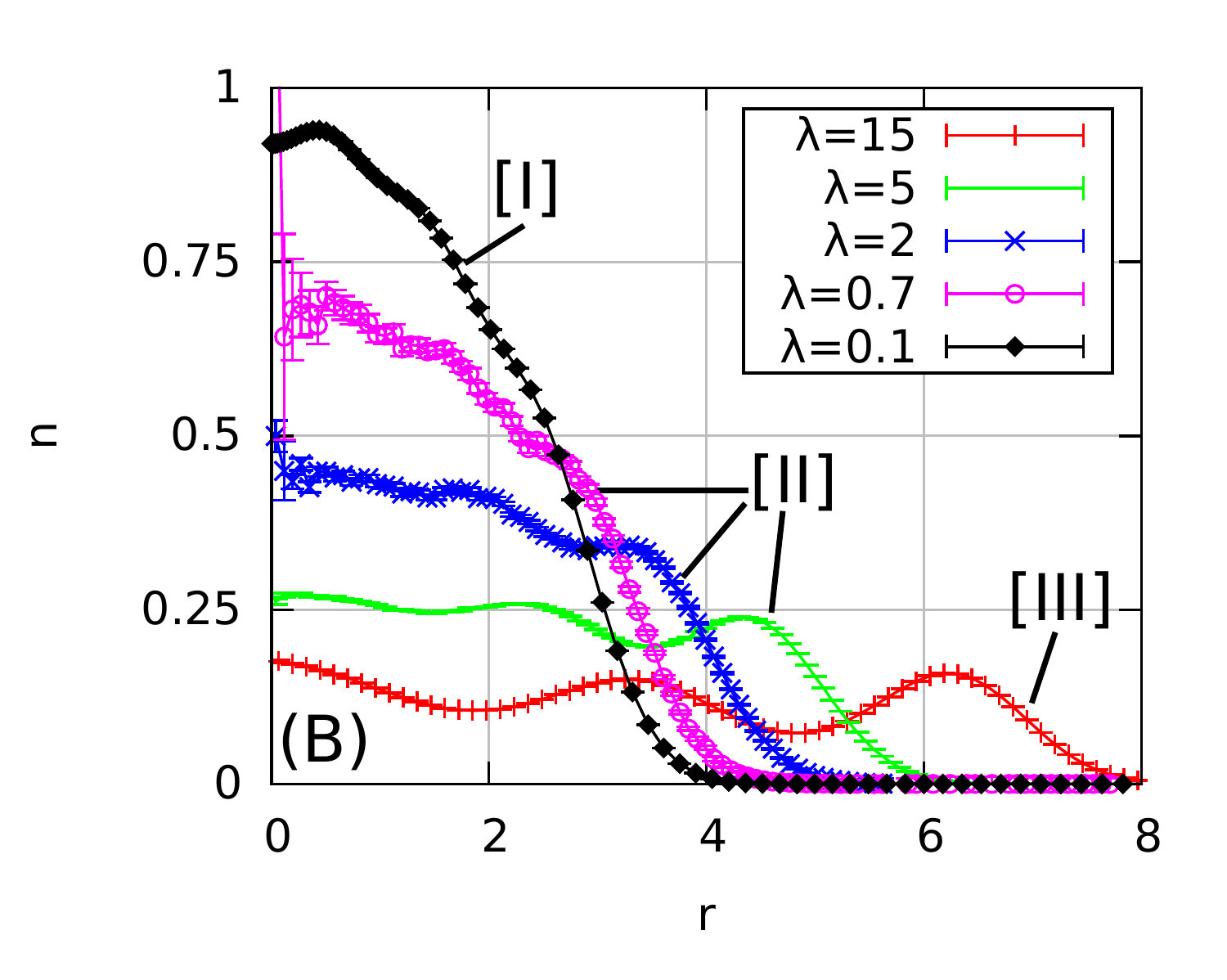}
 \caption{\label{intro}Illustration of the capability of PB-PIMC -- In panel (A), the average sign $S$ from different methods is plotted versus the coupling parameter $\lambda$, Eq.\ (\ref{lambda_def}), for $N=20$ electrons
 in a quantum dot at $\beta=3.0$ (oscillator units). Region [I] denotes the weakly nonideal Fermi gas, [II] the transition region and [III] the strongly correlated regime.
 CPIMC (PIMC) is limited to weak (strong) coupling, i.e.\ to the region left (right) of the blue (green) line.
 Panel (B) shows a comparison of density profiles $n(r)$, plotted versus the distance to the center of the trap $r$, across the entire coupling range.
 }
\end{figure}
Figure \ref{intro} (A) shows the average sign $S$ for $N=20$ electrons, plotted versus the coupling strength $\lambda$, cf.\ Eq.\ (\ref{lambda_def}). CPIMC is applicable in the weakly nonideal regime [I], where 
the system is predominantly shaped by the Fermi statistics. In contrast, standard PIMC allows one to accurately simulate systems in the strongly coupled regime [III], where 
exchange effects are not yet dominating, and bosons and fermions exhibit a very similar behavior.
The PB-PIMC method, as will be shown in this work, is applicable over the entire coupling range yielding reasonably accurate results with acceptable computational effort.
Interestingly, this includes the physically most
interesting transition region [II], where both the Coulomb repulsion and quantum statistics govern the system. Here no ab initio results have been reported to this date,
except for very small particle numbers, since
PIMC and CPIMC fail, due to the sign problem.
In panel (B), we show density profiles from all three regimes.
Evidently, the transition from the strongly coupled system with a pronounced shell structure ($\lambda=15$) to the nearly ideal Fermi gas with the characteristic weak density modulations ($\lambda=0.1$)
can be resolved.

In the remainder of this work, we introduce the PB-PIMC method in detail.
We show that the optimal choice of two free parameters of the fourth-order factorization allows for a calculation of energies
and densities with an accuracy of the order of $0.1\%$ with
as few as two or three propagators, even in the low temperature regime.
We calculate energies and densities from PB-PIMC for $N=20$ electrons at low temperature over the entire coupling range.
We find excellent agreement with both PIMC and CPIMC in the limitting cases of strong and weak coupling, respectively, and perform simulations in the transition regime,
where no other ab initio results are available. Finally, we investigate the performance behavior of our method when the system size is varied.

\section{Theory}\label{theory}
\subsection{Idea of permutation blocking path integral Monte Carlo}

\begin{figure}[]
 \centering
 \includegraphics[width=0.49\textwidth]{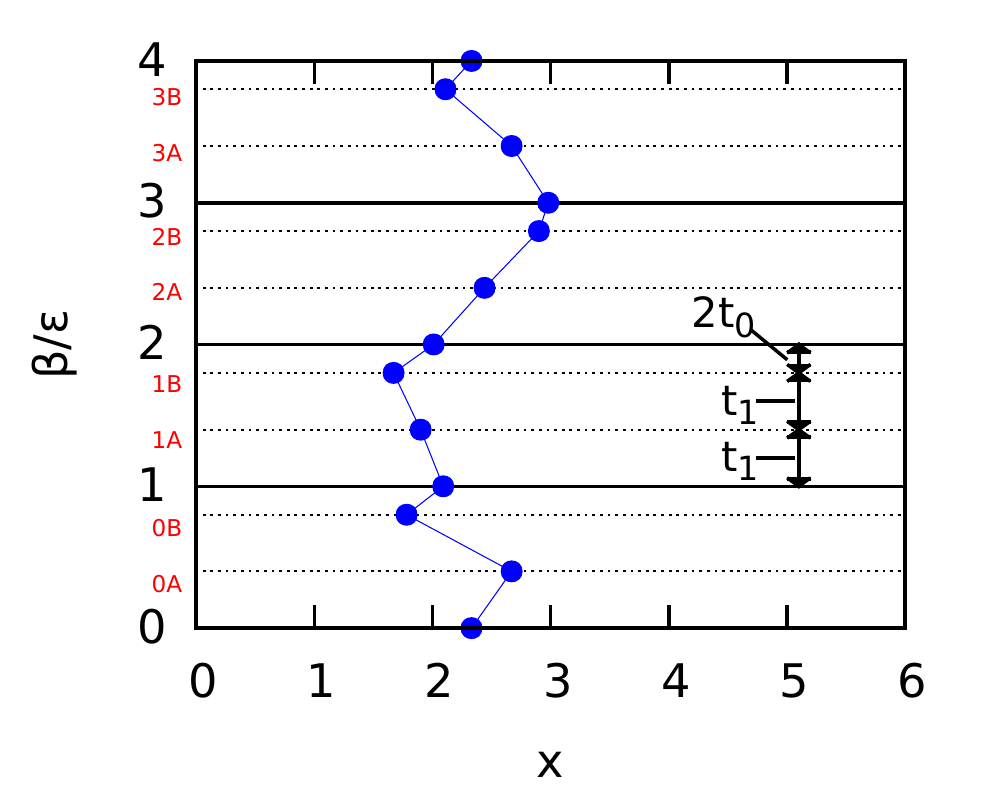}
  \includegraphics[width=0.49\textwidth]{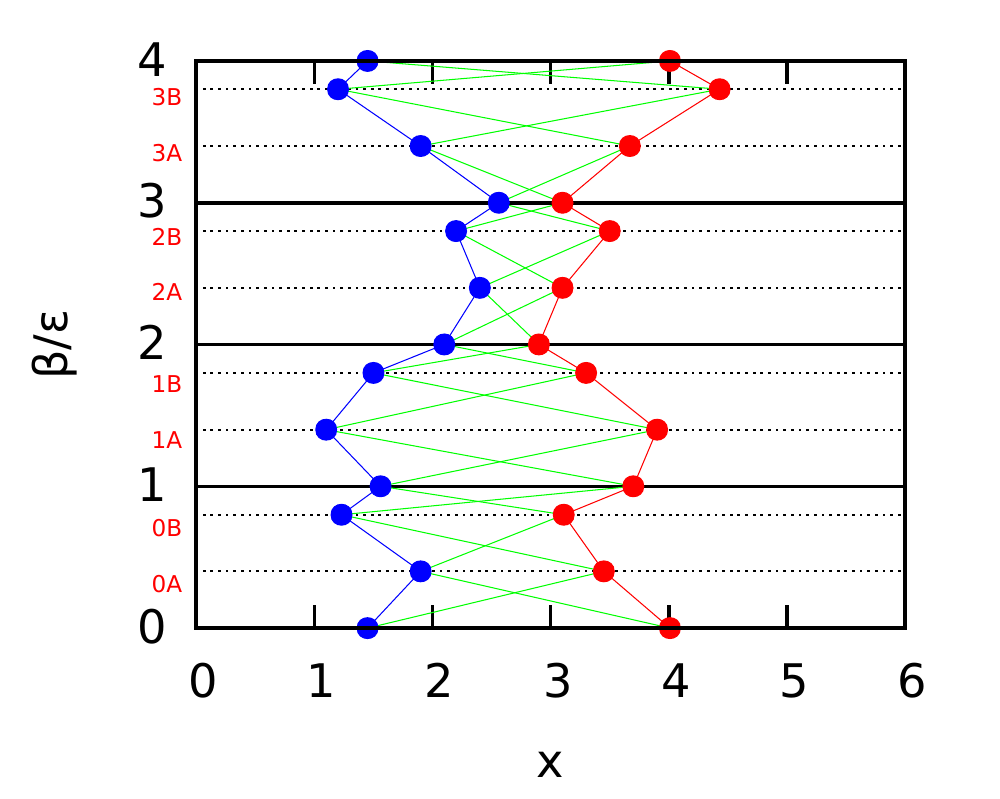}
 \caption{\label{chin}Illustration of the configuration space -- In the left panel, the imaginary time is plotted 
 versus the (arbitrary) spatial coordiante x. Each time step of length $\epsilon$ is further divided into three non-equidistant subintervals, 
 with two ''daughter`` slices $A$ and $B$.
 The right panel illustrates the combination of all $3PN!$ possible trajectories into a single configuration weight $W(\mathbf{X})$.
 Between each two adjacent time slices, both the connection between beads from the same particle (diagonal elements of the diffusion matrix, the blue and red lines)
 and between beads from different particles (off-diagonal elements, the green lines) are efficiently grouped together to improve the average sign.
 }
\end{figure}

We consider the canonical ensemble (the particle number $N$, volume $V$ and inverse temperature $\beta=1/k_\textnormal{B}T$ are fixed) and write the partition function
in coordinate representation as
\begin{eqnarray}
\label{z} Z = \frac{1}{N!} \sum_{\sigma\in S_N} \textnormal{sgn}(\sigma) \int \textnormal{d}\mathbf{R}\ \bra{\mathbf{R}} e^{-\beta\hat{H}} \ket{\hat{\pi}_\sigma\mathbf{R}} \quad ,
\end{eqnarray}
where $\mathbf{R} = \{ \mathbf{r}_1, ... ,\mathbf{r}_N\}$ contains the coordinates of all particles and $\hat{\pi}_\sigma$ denotes the exchange operator 
corresponding to a particular element $\sigma$ from the permutation group $S_N$.
The Hamiltonian is given by the sum of the kinetic ($\hat{K}$) and potential ($\hat{V}$) energy, $\hat{H} = \hat{K} + \hat{V}$.
For the next step, we use the group property of the density operator
\begin{eqnarray}
 \hat{\rho} = e^{-\beta\hat{H}} = \prod_{\alpha=0}^{P-1} e^{-\epsilon\hat{H}}  \quad ,
\end{eqnarray}
with $\epsilon = \beta/P$,
and insert $P-1$ unities of the form $\hat{1} = \int \textnormal{d}\mathbf{R}_\alpha\ \ket{\mathbf{R}_\alpha}\bra{\mathbf{R}_\alpha}$.
This gives
\begin{eqnarray}
 Z =  \int \textnormal{d}\mathbf{R_0}\dots\textnormal{d}\mathbf{R_{P-1}} \prod_{\alpha=0}^{P-1} \left( \frac{1}{N!} \sum_{\sigma\in S_N}\textnormal{sgn}(\sigma) \bra{\mathbf{R}_\alpha} e^{-\epsilon\hat{H}} \ket{\hat{\pi}_{\sigma}\mathbf{R}_{\alpha+1}} \right) \quad . \label{multiprop}
\end{eqnarray}
Note that we have exploited the permutation operator's idempotency property in Eq.\ (\ref{multiprop}) to introduce antisymmetry on all $P$ imaginary time slices.
Following Sakkos \textit{et al.} \cite{sakkos}, we introduce the factorization from \cite{chin2},
\begin{eqnarray}
\label{cchin} e^{-\epsilon\hat{H}} \approx e^{-v_1\epsilon\hat{W}_{a_1}} e^{-t_1\epsilon\hat{K}} e^{-v_2\epsilon\hat{W}_{1-2a_1}} e^{-t_1\epsilon\hat{K}} e^{-v_1\epsilon\hat{W}_{a_1}} e^{-2t_0\epsilon\hat{K}} \quad ,
\end{eqnarray}
for each of the exponential functions in Eq.\ (\ref{multiprop}). By including double commutator terms of the form
\begin{eqnarray}
 [[\hat{V},\hat{K}],\hat{V}] = \frac{\hbar^2}{m} \sum_{i=1}^N |\mathbf{F}_i|^2 \quad ,
\end{eqnarray}
we have to evaluate the total force on each particle, $\mathbf{F}_i = -\nabla_i V(\mathbf{R})$, and Eq.\ (\ref{cchin}) is accurate to fourth order in $\epsilon$.
The explicit form of the modified potential terms $\hat{W}$ is given by
\begin{eqnarray}
 \hat{W}_{a_1} &=& \hat{V} + \frac{u_0}{v_1}a_1\epsilon^2\left( \frac{\hbar^2}{m}\sum_{i=1}^N |\mathbf{F}_i|^2 \right) \quad  \textnormal{and} \\
 \hat{W}_{1-a_1} &=& \hat{V} + \frac{u_0}{v_2}(1-a_1)\epsilon^2\left( \frac{\hbar^2}{m}\sum_{i=1}^N |\mathbf{F}_i|^2 \right) \quad . \nonumber
\end{eqnarray}
There are two free parameters in Eq.\ (\ref{cchin}), namely $0 \le a_1 \le 1$, which controls the relative weight of the forces on a particular slice,
and $0 \le t_0 \le (1-1/\sqrt{3})/2$, which determines the ratio of the, in general, non-equidistant time steps between ''daughter`` slices, cf. Fig.\ \ref{chin}.
All other factors are calculated from these choices:
\begin{eqnarray}
\nonumber u_0 &=& \frac{1}{12}\left( 1 - \frac{1}{1-2t_0} + \frac{1}{6(1-2t_0)^3} \right) \quad , \\
 v_1 &=& \frac{1}{6(1-2t_0)^2} \quad , \\
 v_2 &=& 1-2v_1 \quad \textnormal{and}\nonumber \\
 t_1 &=& \frac{1}{2} - t_0 \quad . \nonumber
\end{eqnarray}
The fourth-order approximation of the imaginary time propagator $e^{-\epsilon\hat{H}}$ is visualized in Fig. \ref{chin}.
The inverse temperature $\beta$ has been split into $P=4$ intervals of length $\epsilon$, which are further divided into three, in general, non-equidistant sub-intervals.
Thus, for each main ''bead`` $\tau_\alpha$, there exist two daughter beads, $\tau_{\alpha A}$ and $\tau_{\alpha B}$.

Let us for a moment ignore the antisymmetry in Eq.\ (\ref{multiprop}) and evaluate the imaginary time propagator in a straightforward way \cite{sakkos}:
\begin{eqnarray}
 \label{chinprop} \bra{\mathbf{R}_\alpha} e^{-\epsilon\hat{H}} \ket{\mathbf{R}_{\alpha+1}} &=&
 \int \textnormal{d}\mathbf{R}_{\alpha A}\textnormal{d}\mathbf{R}_{\alpha B}\ \left[
 e^{-\epsilon\tilde V_\alpha} e^{-u_0\epsilon^3 \frac{\hbar^2}{m} \tilde F_\alpha}  \right. \\
 \nonumber & & \left. \prod_{i=1}^N \rho_\alpha(i,i)\rho_{\alpha,A}(i,i) \rho_{\alpha B}(i,i) \right] \quad ,
 \end{eqnarray}
with the definitions of the potential terms
\begin{eqnarray}
 \tilde V_\alpha &=& v_1 V(\mathbf{R}_\alpha) + v_2 V(\mathbf{R}_{\alpha A}) + v_1 V(\mathbf{R}_{\alpha B}) \quad ,\\
 \nonumber \tilde F_\alpha &=& \sum_{i=1}^N \left( a_1 |\mathbf{F}_{\alpha,i}|^2 + (1-2a_1) |\mathbf{F}_{\alpha A,i}|^2 + a_1 |\mathbf{F}_{\alpha B,i}|^2 \right)  \quad ,
\end{eqnarray}
and the diffusion matrices
\begin{eqnarray}
 \rho_\alpha(i,j) &=& \lambda_{t_1\epsilon}^{-D} \textnormal{exp} \left( -\frac{\pi}{\lambda^2_{t_1\epsilon}} ( \mathbf{r}_{\alpha,j} - \mathbf{r}_{\alpha A,i})^2 \right) \quad , \nonumber \\
  \label{dm} \rho_{\alpha A}(i,j) &=& \lambda_{t_1\epsilon}^{-D} \textnormal{exp} \left( -\frac{\pi}{\lambda^2_{t_1\epsilon}} ( \mathbf{r}_{\alpha A,j} - \mathbf{r}_{\alpha B,i})^2 \right) \quad , \\
 \nonumber \rho_{\alpha B}(i,j) &=& \lambda_{2t_0\epsilon}^{-D} \textnormal{exp} \left( -\frac{\pi}{\lambda^2_{2t_0\epsilon}} ( \mathbf{r}_{\alpha B,j} - \mathbf{r}_{\alpha+1,i})^2 \right) \quad ,
\end{eqnarray}
 where $\lambda_\beta$ denotes the thermal wavelength $\lambda_\beta^2 = 2\pi \hbar^2 \beta / m$ and $D$ is the dimensionality of the system.
 Thus, the matrix elements of Eq.\ (\ref{dm}) are equal to the free particle density matrix, $\rho_\alpha(i,j) = \rho_0(\mathbf{r}_{\alpha,j},\mathbf{r}_{\alpha A,i},t_1\epsilon)$.
The permutation operator commutes with both $\hat{\rho}$ and $\hat{H}$ and we are, therefore, 
allowed to artificially introduce the antisymmetrization between all $3P$ slices without changing the result.
This transforms Eq.\ (\ref{chinprop}) to
\begin{eqnarray}
 \nonumber\frac{1}{N!} \sum_{\sigma\in S_N}\textnormal{sgn}(\sigma) \bra{\mathbf{R}_\alpha} e^{-\epsilon\hat{H}} \ket{\hat{\pi}_\sigma\mathbf{R}_{\alpha+1}} =
 \left(\frac{1}{N!}\right)^3 \int \textnormal{d}\mathbf{R}_{\alpha A}\textnormal{d}\mathbf{R}_{\alpha B}\ 
\\  \left[  e^{-\epsilon\tilde V_\alpha}    e^{-\epsilon^3 u_0 \frac{\hbar^2}{m} \tilde F_\alpha} \textnormal{det}(\rho_\alpha)\textnormal{det}(\rho_{\alpha A})\textnormal{det}(\rho_{\alpha B}) \right] \quad .
\end{eqnarray}
Finally, this gives the partition function
\begin{eqnarray}
\label{finalz} Z = \frac{1}{(N!)^{3P}} \int \textnormal{d}\mathbf{X} \prod_{\alpha=0}^{P-1} e^{-\epsilon\tilde V_\alpha}e^{-\epsilon^3u_0\frac{\hbar^2}{m}\tilde F_\alpha}\textnormal{det}(\rho_\alpha)\textnormal{det}(\rho_{\alpha A})\textnormal{det}(\rho_{\alpha B}) \quad ,
\end{eqnarray}
and the integration is carried out over all coordinates on all $3P$ slices:
\begin{eqnarray}
 \label{bz}\textnormal{d}\mathbf{X} = \textnormal{d}\mathbf{R}_0\dots\textnormal{d}\mathbf{R}_{P-1}\textnormal{d}\mathbf{R}_{0A}\dots \textnormal{d}\mathbf{R}_{P-1A}\textnormal{d}\mathbf{R}_{0B}\dots \textnormal{d}\mathbf{R}_{P-1B} \quad .
\end{eqnarray}

The benefits of the partition function Eq.\ (\ref{finalz}) are illustrated in the right panel of Fig.\ \ref{chin} where the beads of two particles are 
plotted in the $\tau$-$x$-plane.
In the usual PIMC formulation (without the determinants), each of the particles would correspond to a single closed trajectory as visualized by
the blue and red connections. To take into account the antisymmetry of fermions, one would also need to sample all configurations with the 
same positions of the individual beads but different connections between adjacent time slices, which have both positive and negative weights.
By indroducing determinants between all slices, we include all $N!$ possible connections between beads on adjacent slices (the green lines) into a 
single configuration weight and the usual interpretation of mapping a quantum system onto an ensemble of interacting ringpolymers \cite{chandler} is no longer appropriate.
Therefore, a large number of sign changes, due to different permutations, are grouped together resulting in an efficient compensation of many terms (blocking),
and the average sign (cf.\ Eq.\ (\ref{asign})) in our simulations is significantly increased \cite{lyu}.

\subsection{Energy estimator}
The total energy $E$ follows from the partition function via the familiar relation
\begin{eqnarray}
 E = - \frac{1}{Z} \frac{\partial Z}{\partial \beta} \quad . \label{tde}
\end{eqnarray}
Substituting the expression from Eq.\ (\ref{finalz}) into (\ref{tde}) and performing a lengthy but straightforward calculation gives the final result for the thermodynamic (TD) estimator
\begin{eqnarray}
 \nonumber E &=& \frac{3DN}{2\epsilon} - \sum_{k=0}^{P-1}\sum_{\kappa=1}^N\sum_{\xi=1}^N \left(
 \frac{\pi \Psi^{k}_{\kappa\xi} }{\epsilon P \lambda^2_{t_1\epsilon}} (\mathbf{r}_{k,\kappa} - \mathbf{r}_{kA,\xi} )^2 \right.
 \\ \label{energy} & & + \left. \frac{\pi \Psi^{kA}_{\kappa\xi}}{\epsilon P \lambda^2_{t_1\epsilon}} (\mathbf{r}_{kA,\kappa} - \mathbf{r}_{kB,\xi} )^2
 +  \frac{\pi \Psi^{kB}_{\kappa\xi}}{\epsilon P \lambda^2_{2t_0\epsilon}} (\mathbf{r}_{kB,\kappa} - \mathbf{r}_{k+1,\xi} )^2
 \right) \\ \nonumber & & + \frac{1}{P} \sum_{k=0}^{P-1} \left( \tilde V_k + 3 \epsilon^2 u_0 \frac{\hbar^2}{m} \tilde F_k \right) \quad ,
\end{eqnarray}
with the definitions
\begin{eqnarray}
 \Psi^{k}_{\kappa\xi} &=& \left( \rho^{-1}_{k} \right)_{\kappa\xi} \left(\rho_{k}\right)_{\xi\kappa} \\
  \Psi^{kA}_{\kappa\xi} &=& \left( \rho^{-1}_{kA} \right)_{\kappa\xi} \left(\rho_{kA}\right)_{\xi\kappa} \nonumber  \\ \nonumber
   \Psi^{kB}_{\kappa\xi} &=& \left( \rho^{-1}_{kB} \right)_{\kappa\xi} \left(\rho_{kB}\right)_{\xi\kappa}  \quad .
\end{eqnarray}
To split the total energy into a kinetic and a potential part, we evaluate
\begin{eqnarray}
 K = \frac{m}{\beta Z} \frac{\partial}{\partial m} Z \quad , \label{uz}
\end{eqnarray}
and find the TD estimator of the kinetic energy
\begin{eqnarray}
 K &=& \frac{3ND}{2\epsilon} - \sum_{k=0}^{P-1}\sum_{\kappa=1}^N\sum_{\xi=1}^N \left[
 \frac{ \pi \Psi^{k}_{\kappa\xi} }{ \epsilon P \lambda_{t_1\epsilon}^2 } (\mathbf{r}_{k,\kappa} - \mathbf{r}_{kA,\xi})^2 
 +\frac{ \pi \Psi^{kA}_{\kappa\xi} }{ \epsilon P \lambda_{t_1\epsilon}^2 } (\mathbf{r}_{kA,\kappa} - \mathbf{r}_{kB,\xi})^2 
\right. \nonumber \\  & & + \left. \frac{ \pi \Psi^{kB}_{\kappa\xi} }{ \epsilon P \lambda_{2t_0\epsilon}^2 } (\mathbf{r}_{kB,\kappa} - \mathbf{r}_{k+1,\xi})^2 
 \right] + \frac{1}{P} \sum_{k=0}^{P-1} \left( \epsilon^2 u_0 \frac{ \hbar^2}{m} \tilde F_k \right) \quad .
\end{eqnarray}
Thus, the estimator of the potential energy is given by
\begin{eqnarray}
 V = E - K = \frac{1}{P} \sum_{k=0}^{P-1} \left( \tilde V_k + 2\epsilon^2 u_0 \frac{\hbar^2}{m} \tilde F_k \right) \quad .
\end{eqnarray}
We notice that the forces contribute to both the kinetic and the potential energy.
For completeness, we mention that, for an increasing number of propagators, $P\to\infty$, the first and second terms in Eq.\ (\ref{energy}) diverge, which leads to
a growing variance and, therefore, statistical uncertainty of both $E$ and $K$. To avoid this problem, one might derive a virial estimator, e.g.\ \cite{energy}, which
requires the evaluation of the derivative of the potential terms instead. However, since we are explicitly interested in performing simulations with few propagators to
relieve the fermion sign problem, the estimator from Eq.\ (\ref{energy}) is sufficient.

\section{Monte Carlo algorithm}
In section \ref{theory}, we have derived an expression for the partition function $Z$, Eq.\ (\ref{finalz}), which incorporates determinants of the diffusion matrices between all $3P$ time slices,
thereby combining $3PN!$ different configurations from the usual PIMC into a single weight $W(\mathbf{X})$. However, each determinant can still be either positive or negative, depending
on the relative magnitude of diagonal and off-diagonal elements. Hence, we apply the Metropolis algorithm \cite{metropolis} to the modified partition function
\begin{eqnarray}
 Z^{'} = \int \textnormal{d}\mathbf{X}\ |W(\mathbf{X})| \label{priam} \quad ,
\end{eqnarray}
and calculate fermionic expectation values as
\begin{eqnarray}
 \braket{O}_\textnormal{f} = \frac{ \braket{ OS }^{'} }{ \braket{S}^{'} } \quad ,
\end{eqnarray}
with the definition of the average sign
\begin{eqnarray}
 \label{asign} \braket{S}^{'} = \frac{1}{Z^{'}} \int \textnormal{d}\mathbf{X}\ |W(\mathbf{X})| S(\mathbf{X}) \quad ,
\end{eqnarray}
and the signum of the configuration $\mathbf{X}$,
\begin{eqnarray}
 S(\mathbf{X}) = \prod_{\alpha=0}^{P-1} \left[ \textnormal{sgn}(\textnormal{det}(\rho_\alpha))
 \textnormal{sgn}(\textnormal{det}(\rho_{\alpha A})) \textnormal{sgn}(\textnormal{det}(\rho_{\alpha B})) \right] \quad .
\end{eqnarray}
\begin{figure}[]
 \centering
 \includegraphics[width=0.49\textwidth]{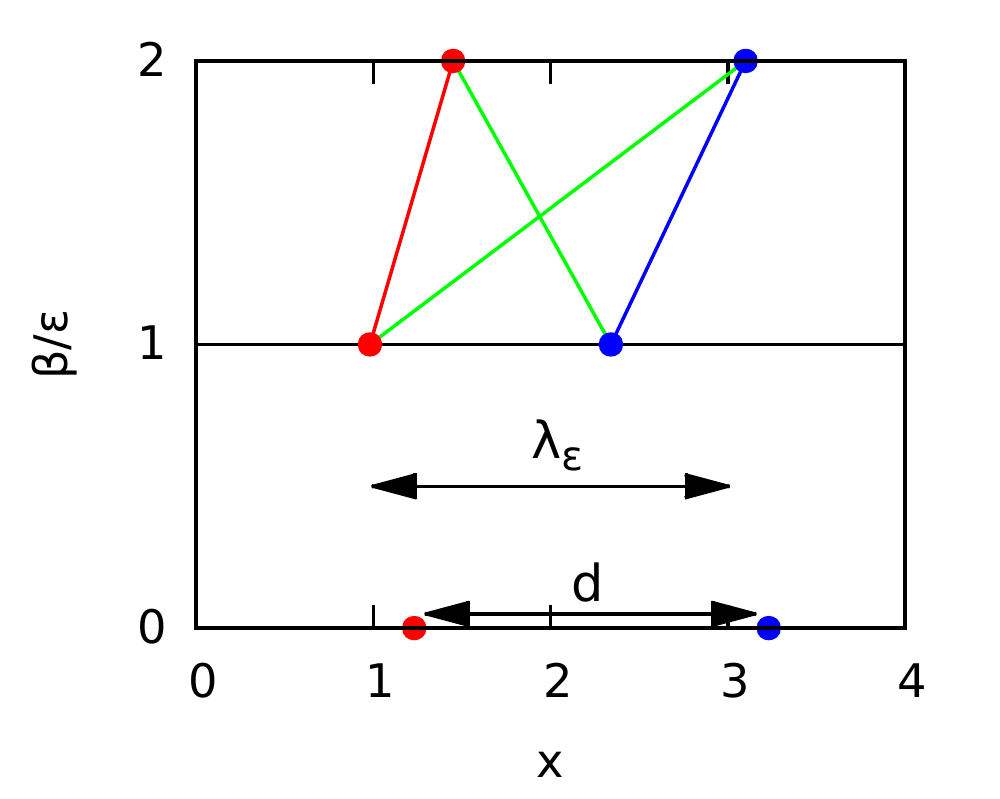}
  \includegraphics[width=0.49\textwidth]{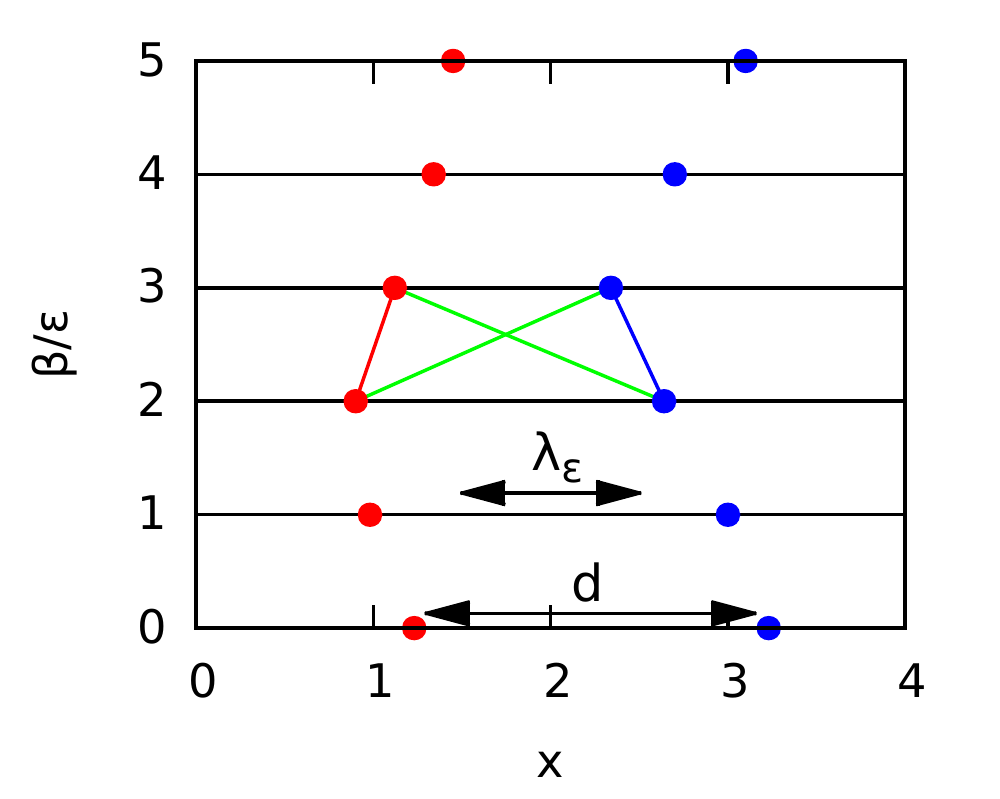}
 \caption{\label{lambda}Influence of the imaginary time step $\epsilon$ on the efficiency of the permutation blocking --
 Two configurations of $N=2$ particles are visualized in the $\tau$-$x$-plane. In the left and right panel, there are $P=2$ and $P=5$ time slices, respectively (daughter slices are neglected for simplicity).
 Only with few propagators, the thermal wavelength $\lambda_\epsilon$ of a single propagator is comparable to the mean interparticle distance $d$, which
 is crucial for an efficient grouping of permutations into a single configuration weight.
 With increasing $P$, diagonal (red and blue lines) and off-diagonal (green lines) distances are no longer of the same order and the permutation blocking
 is inefficient.
 }
\end{figure}
Let us summarize some important facts about the configuration space defined by Eq.\ (\ref{priam}):
\begin{enumerate}
 \item With increasing number of propagators $P$, the effect of the blocking decreases and, for $P\to\infty$, the sign converges to the sign of standard PIMC.
 Blocking is maximal if $\lambda_{t_1\epsilon}$ and $\lambda_{2t_0\epsilon}$ are comparable to the average interparticle distance $d$, cf. Fig.\ \ref{lambda}.
 Only in such a case, there can be both large diagonal and off-diagonal elements in the diffusion matrices.
\item Configuration weights $|W(\mathbf{X})|$ can only be large, when at least one element in each row of each diffusion matrix is large. Therefore,
we sample either large diagonal or large off-diagonal elements. Blocking happens naturally as a by-product and does not have to be specifically included into the sampling.
This also means that we have to implement a mechanism to sample exchange, i.e., to switch between large diagonal and off-diagonal diffusion matrix elements.
\item There are no fixed trajectories. Therefore, beads do not have a previous or a next bead, as in standard PIMC. For an efficient and flexible sampling algorithm, we temporarily construct
artificial trajectories and choose the included beads randomly.
 \end{enumerate}
The most efficient mechanism for the sampling of exchange cycles in standard PIMC is the so-called worm algorithm \cite{bon,bon2}, where macroscopic trajectories are
naturally realized by a small set of local updates which enjoy a high acceptance probability. In the rest of the section, we
modify this algorithm to be applicable to the new configuration space without any fixed connections between individual beads.

 \subsection{Sampling scheme}

 \begin{figure}[]
 \centering
 \includegraphics[width=0.49\textwidth]{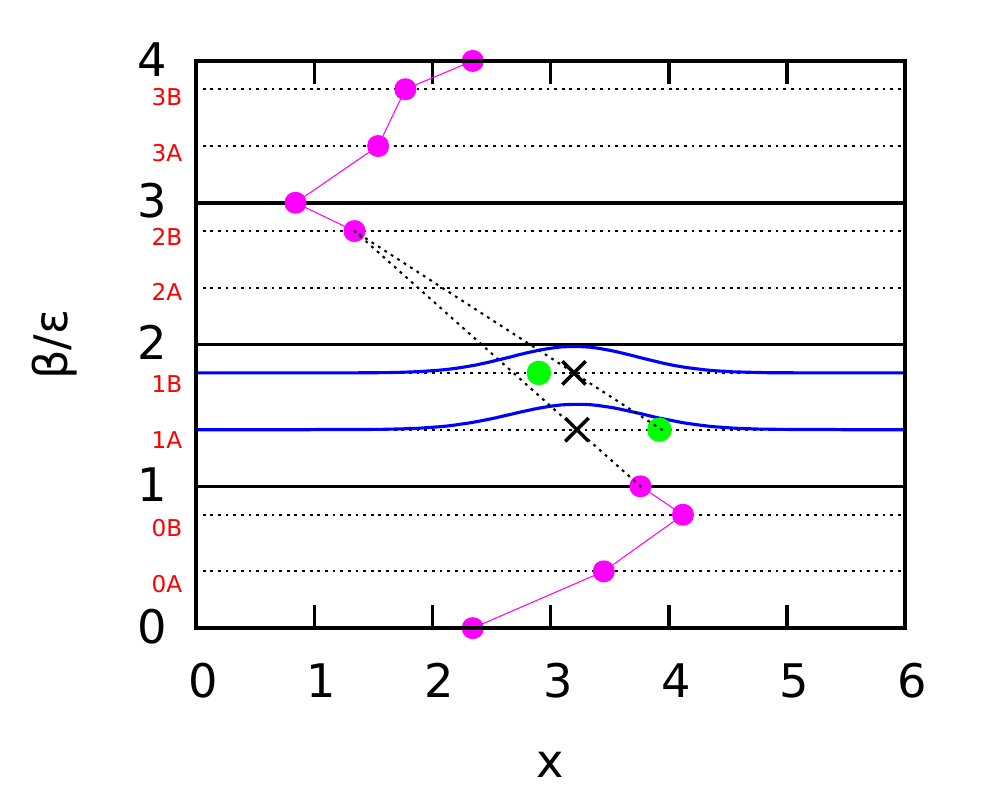}
  \includegraphics[width=0.49\textwidth]{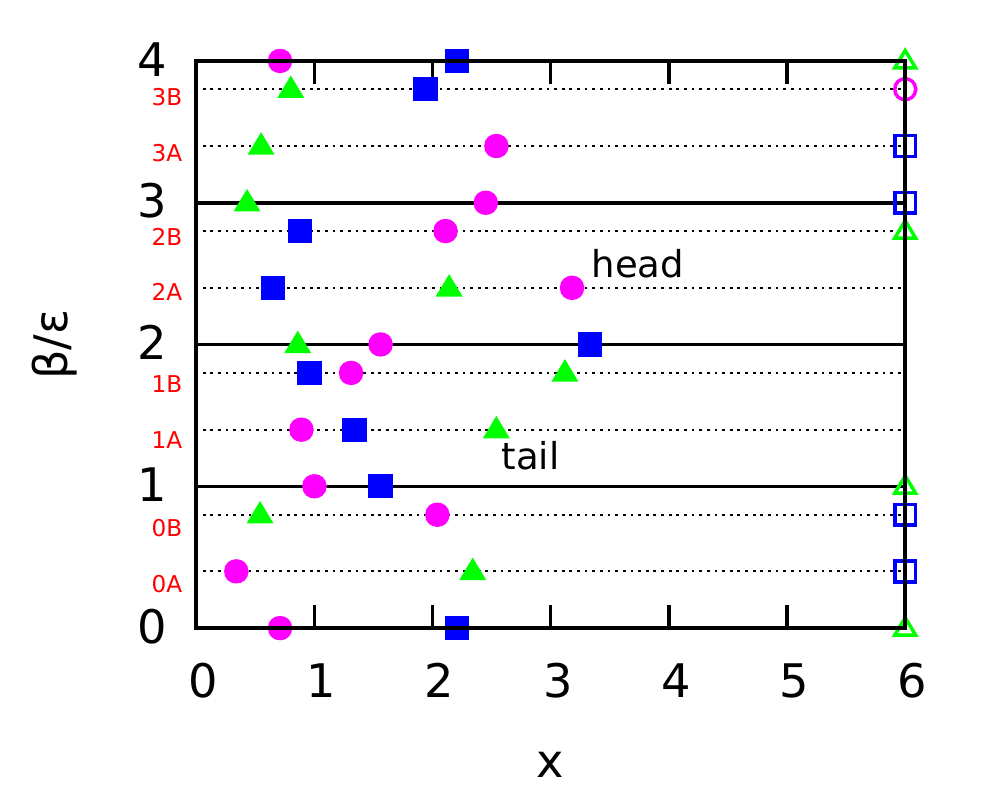}
 \caption{\label{sampling}Illustration of the sampling scheme (left) and the extended configuration space (right) -- In the left panel, an artificial trajectory (pink curve) with four missing beads is plotted in the $\tau$-$x$-plane.
 The new coordinates (green circles) are sampled according to a Gaussian (blue curves) around the intersection of the connecting straight lines between the previous and last bead with the current time slice (black crosses).
 The right panel gives an example for an open configuration in the extended configuration space with two special beads which are denoted as ''head`` and ''tail``.
 There are only $N-1$ beads on eight time slices, going forward in imaginary time 
 starting from $\tau_\textnormal{head} = \tau_{2A}$.
 The circles, triangles and squares distinguish beads from three different particles and the empty symbols at the right boundary indicate the missing beads on a particular slice.}
\end{figure}

To take advantage of the main benefits from the usual continuous space worm algorithm, we will temporarily construct artificial trajectories and sample new beads according
to standard PIMC techniques, e.g. \cite{charge}.
The initial situation for our considerations is illustrated in the left panel of Fig.\ \ref{sampling}, where a pre-existing trajectory (pink curve) with four missing beads in the middle is shown in the $\tau$-$x$-plane.
We choose the sampling probability to close the configuration as
\begin{eqnarray}
 \label{connect} T_\textnormal{sample} = \frac{\prod_{i=0}^{M-1} \rho_0(\mathbf{r}_i, \mathbf{r}_{i+1}, \tau_{i+1} - \tau_i)}{\rho_0(\mathbf{r}_0, \mathbf{r}_M, \tau_M - \tau_0) }  \quad ,
\end{eqnarray}
which results in the consecutive generation of $M-1$ new coordinates $\mathbf{r}_i$, $i\in[1,M-1]$, according to
\begin{eqnarray}
 \label{acc}P(\mathbf{r}_i) &=& \frac{ \rho_0(\mathbf{r}_{i-1}, \mathbf{r}_i, \tau_i - \tau_{i-1}) \rho_0(\mathbf{r}_i, \mathbf{r}_M, \tau_M- \tau_i) }{ \rho_0(\mathbf{r}_{i-1}, \mathbf{r}_{M}, \tau_M-\tau_{i-1}) } \\ \nonumber
 &=& \left( \frac{1}{ \sqrt{ 2\pi \sigma_i^2 } } \right)^D \textnormal{exp}\left( - \frac{ (\mathbf{r}_i - \mathbf{\xi}_i)^2 }{ 2\sigma_i^2 } \right) \quad ,
\end{eqnarray}
which is a Gaussian (cf. the blue curves in Fig.\ \ref{sampling}) with the variance
\begin{eqnarray}
 \sigma_i^2 =\frac{\hbar^2}{m} \frac{ (\tau_i - \tau_{i-1}) (\tau_M - \tau_i) }{ \tau_M - \tau_{i-1} } \quad ,
\end{eqnarray}
around the intersection of the connection between the previous coordinate, $\mathbf{r}_{i-1}$, with the end point $\mathbf{r}_M$ and the time slice $\tau_i$
\begin{eqnarray}
 \mathbf{\xi}_i = \frac{ \tau_M - \tau_i}{\tau_M - \tau_{i-1}} \mathbf{r}_{i-1} + \frac{\tau_i - \tau_{i-1}}{\tau_M - \tau_{i-1}} \mathbf{r}_M \quad .
\end{eqnarray}

 \subsection{Artificial worm algorithm}\label{updates}
 In the usual WA-PIMC, the configuration space is defined by the Matsubara Green function (MGF, e.g. \cite{walecka}) which implies that the algorithm
 does not only allow for the change of the particle number $N$ (grand canonical ensemble) but, in addition, requires the generation of configurations
 with a single open path, the so-called worm.
 However, in the PB-PIMC configuration space defined by Eq.\ (\ref{finalz}), there are no trajectories and, therefore, no direct realization of a worm is possible.
 Instead, we consider an extended ensemble, which combines closed configurations with a total of $3NP$ beads and open configurations, where on some consecutive time slices
 the number of beads is reduced by one, to $N-1$.
Such a configuration is illustrated in the right panel of Fig.\ \ref{sampling}.
There are two special beads which are denoted as ''head`` and ''tail`` and the triangles, circles and squares symbolize beads from three different particles.
There are eight beads from different particles missing (indicated by the empty symbols at the right boundary) between $\tau_\textnormal{head} = \tau_{2A}$ and $\tau_\textnormal{tail} = \tau_{1A}$,
going forward in imaginary time.

 For most slices, the computation of the diffusion matrix allows for no degree of freedom in the extended ensemble. 
We define the latter in a way, that the head bead does not serve as a starting point for the elements but is treated as if it was missing.
This is justified because, otherwise, there does not necessarily exist a large matrix element in this particular row because no artificial connection has been sampled on the next slice.
For the configuration from Fig. \ref{sampling}, the diffusion matrix of the head's time slice is given by
\begin{eqnarray}
\rho_{2A} &=& 
 \begin{pmatrix}
\rho_0(\mathbf{r}_{1,2A}, \mathbf{r}_{1,2B}, t_1\epsilon) & \rho_0(\mathbf{r}_{1,2A}, \mathbf{r}_{2,2B}, t_1\epsilon) & 0\\
1 & 1 & 1 \\
\rho_0(\mathbf{r}_{3,2A}, \mathbf{r}_{1,2B}, t_1\epsilon) & \rho_0(\mathbf{r}_{3,2A}, \mathbf{r}_{2,2B}, t_1\epsilon) & 0
\end{pmatrix} \\  \Rightarrow
\textnormal{det}(\rho_{2A}) &=& \textnormal{det}
 \begin{pmatrix}
\rho_0(\mathbf{r}_{1,2A}, \mathbf{r}_{1,2B}, t_1\epsilon) & \rho_0(\mathbf{r}_{1,2A}, \mathbf{r}_{2,2B}, t_1\epsilon)\\
\rho_0(\mathbf{r}_{3,2A}, \mathbf{r}_{1,2B}, t_1\epsilon) & \rho_0(\mathbf{r}_{3,2A}, \mathbf{r}_{2,2B}, t_1\epsilon) \\
\end{pmatrix} \quad . \nonumber 
\end{eqnarray}
All diffusion matrices with $N-1$ beads on their slices are computed in the same way.
The other degree of freedom for which the extended ensemble allows is the choice whether the tail will be included as the final coordinate in the diffusion matrix or not.
Here, it makes sense to allow for this possibility, because there does exist at least a single large element in this particular row anyway.
The corresponding matrix for the configuration from Fig. \ref{sampling} looks like
\begin{eqnarray}
\rho_{2A} &=& 
 \begin{pmatrix}
\rho_0(\mathbf{r}_{1,1}, \mathbf{r}_{1,1A}, t_1\epsilon) & \rho_0(\mathbf{r}_{1,1}, \mathbf{r}_{2,1A}, t_1\epsilon) & \rho_0(\mathbf{r}_{1,1}, \mathbf{r}_{3,1A}, t_1\epsilon)\\
\rho_0(\mathbf{r}_{2,1}, \mathbf{r}_{1,1A}, t_1\epsilon) & \rho_0(\mathbf{r}_{2,1}, \mathbf{r}_{2,1A}, t_1\epsilon) & \rho_0(\mathbf{r}_{2,1}, \mathbf{r}_{3,1A}, t_1\epsilon) \\
1 & 1 & 1 
\end{pmatrix} \quad .
\end{eqnarray}
However, we emphasize that the particular choice of the extended ensemble does not influence the extracted canonical expectation values as long as detailed balance is
fulfilled in all updates.
We have developed a simulation scheme which consists of four different types of moves that ensure detailed balance and ergodicity.
The updates are presented in detail in \ref{app}.

\section{Simulation results}
As a test system to benchmark our method, we consider $N$ spin-polarized electrons in a quantum dot \cite{egger,wigner,rontani,ghosal}, which can be described approximately by a harmonic confinement with a frequency $\Omega$.
We use oscillator units, i.e., the characteristic energy scale $E_0 = \hbar\Omega$ and oscillator length $l = \sqrt{\hbar/\Omega m}$, and obtain the dimensionless Hamiltonian
\begin{eqnarray}
 \label{hamilton}\hat{H} = -\frac{1}{2}\sum_{i=1}^N \nabla_i^2 + \frac{1}{2} \sum_{i=1}^N \mathbf{r}_i^2 +  \sum_{i<j}^N \frac{ \lambda }{ |\mathbf{r}_i - \mathbf{r}_j|} \quad ,
\end{eqnarray}
with the coupling parameter 
\begin{eqnarray}
 \lambda = \frac{ e^2 }{l_0  \hbar \Omega} \quad , \label{lambda_def}
\end{eqnarray}
being defined as the ratio of Coulomb and oscillator energy.
For large $\lambda$, the electrons are strongly coupled and exchange effects become negligible (region [III] in Fig.\ \ref{intro}), while, for $\lambda\ll1$,
the ideal Fermi gas will be approached and the system is governed by the fermionic exchange (region [I] in Fig.\ \ref{intro}).
To confirm the quality of our simulations, we compare the results at weak and strong coupling with CPIMC and standard PIMC, respectively, where they are available.

\subsection{Optimal choice of $a_1$ and $t_0$}
We start the discussion of the simulation results by investigating the effects of the two free parameters $a_1$ and $t_0$ on the convergence of 
two different observables, namely the energy $E$ and radial density $n(r)$.
  \begin{figure}[]
 \centering
 \includegraphics[width=0.49\textwidth]{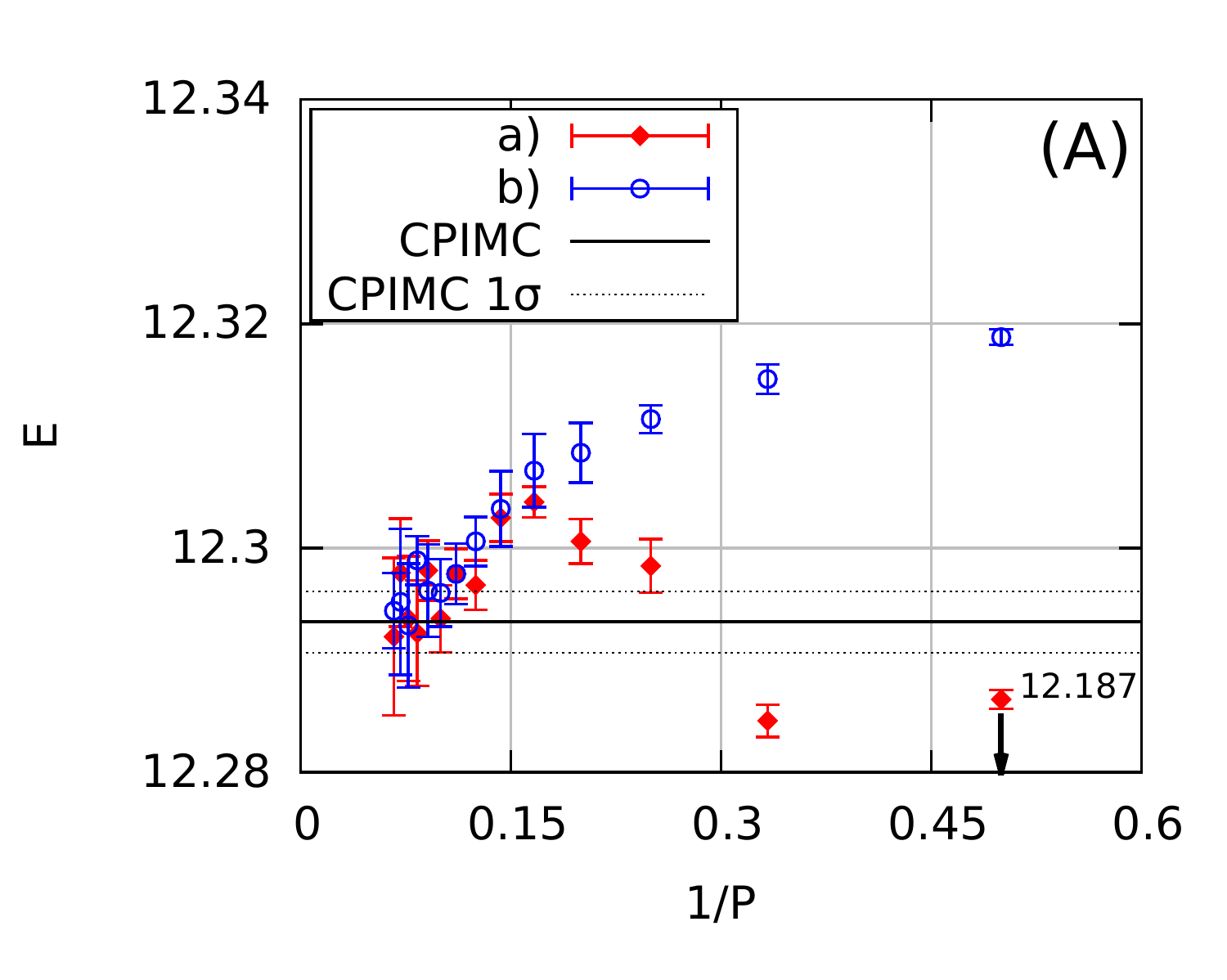}
  \includegraphics[width=0.49\textwidth]{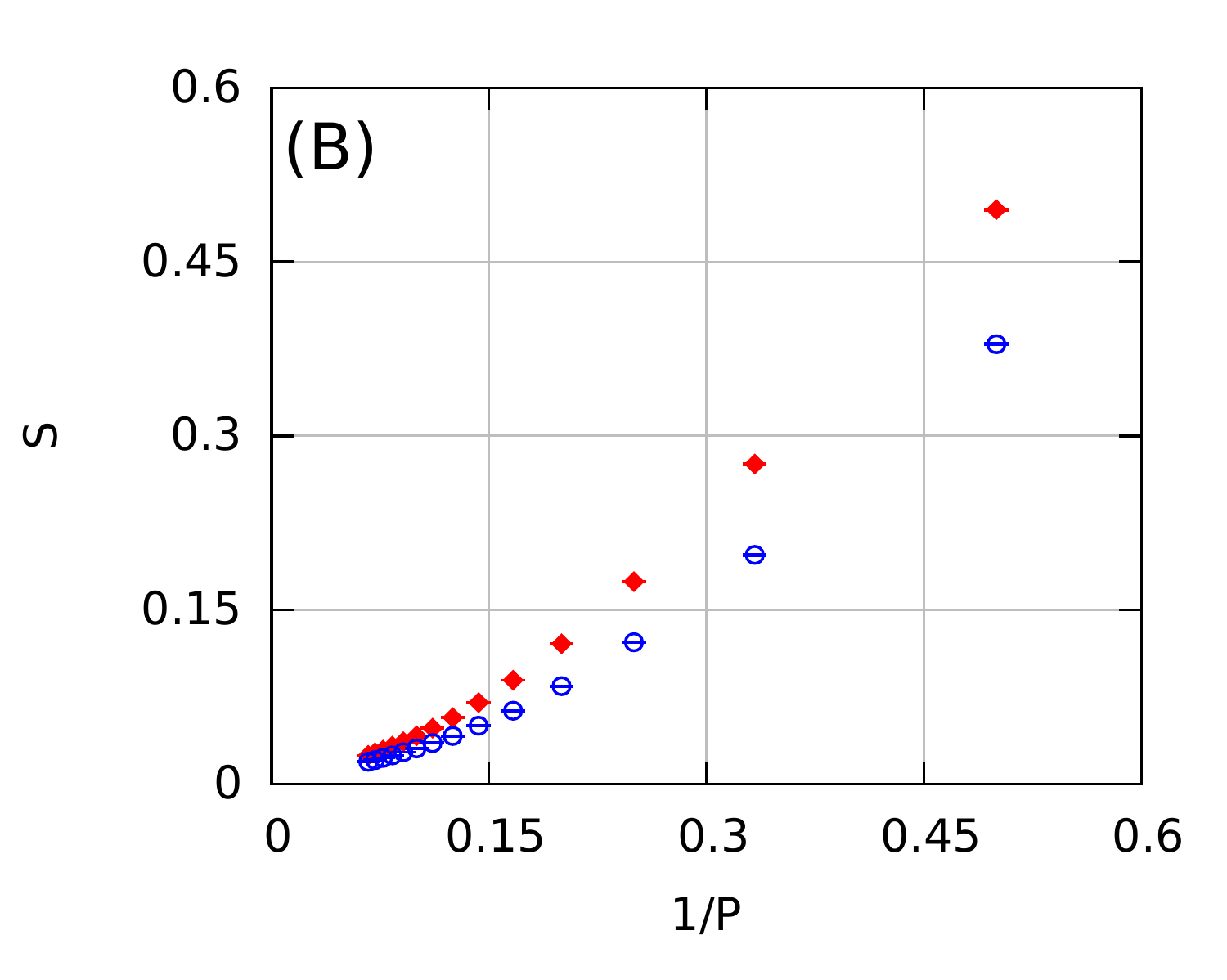}\\
  \includegraphics[width=0.49\textwidth]{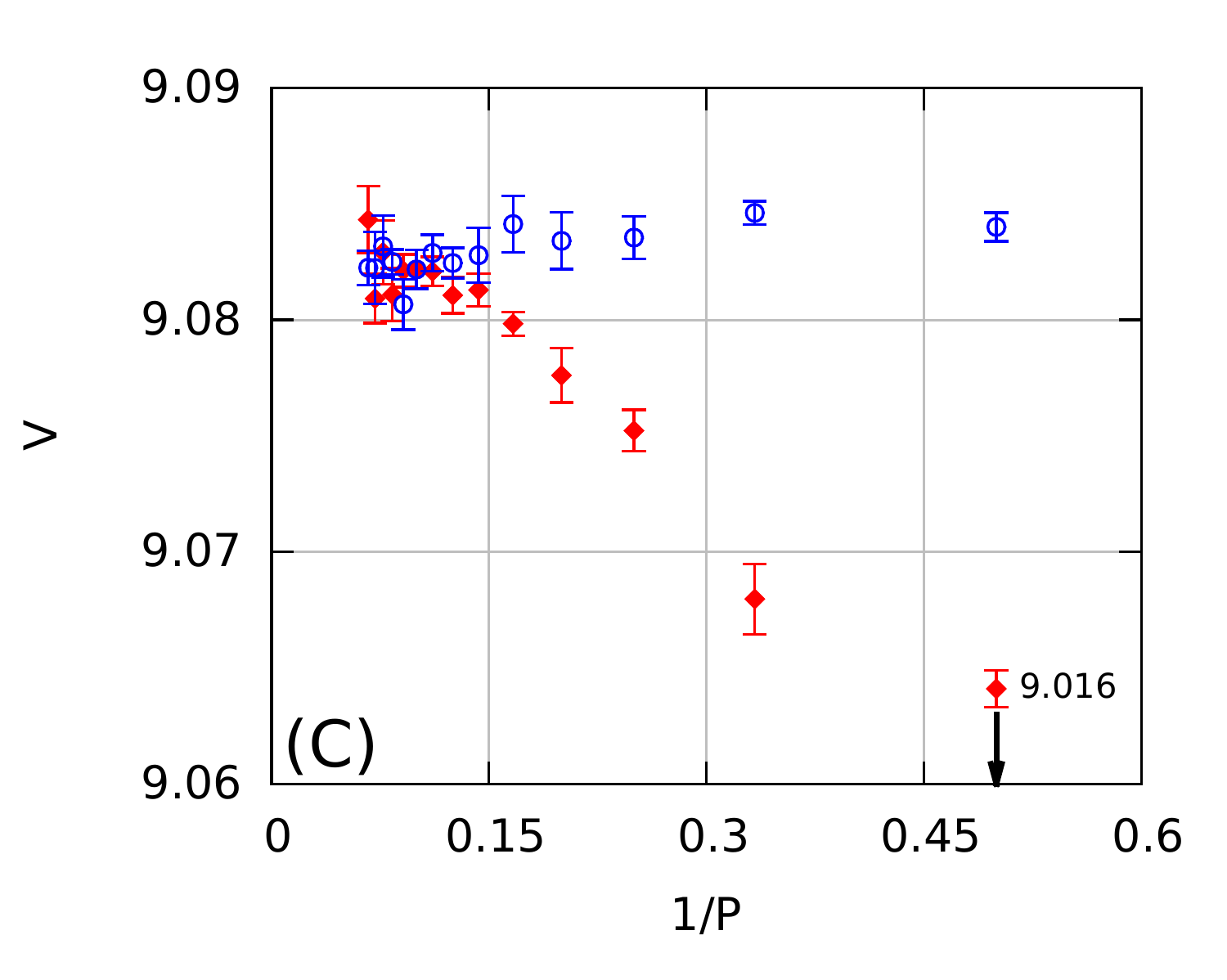}
  \includegraphics[width=0.49\textwidth]{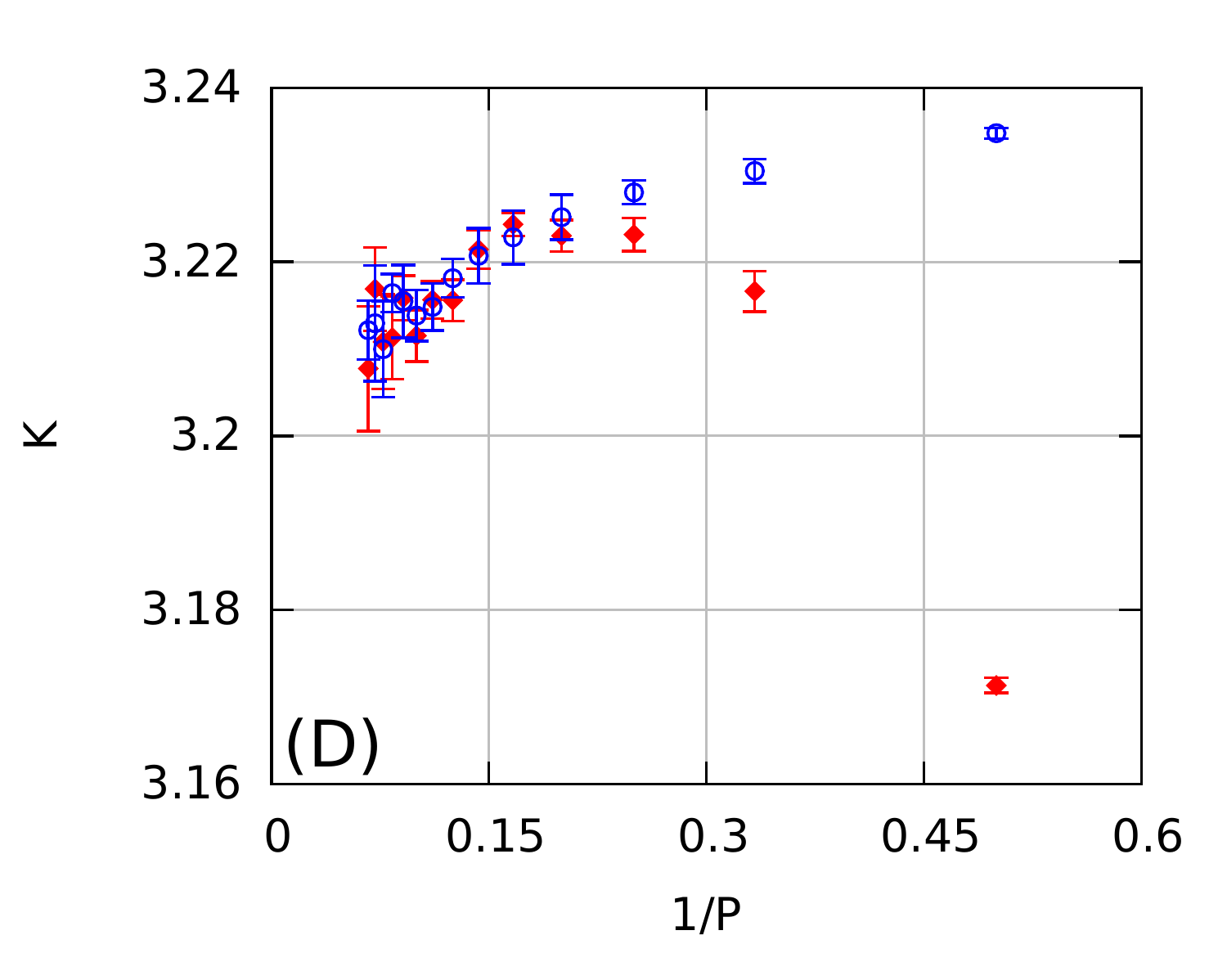}
 \caption{\label{conv}Convergence of the energy for $N=4$, $\lambda=1.3$ and $\beta=5.0$ -- 
 Panel (A) shows the convergence of the total energy versus the inverse number of propagators $P^{-1} \propto \epsilon$.
 Shown are the results for two different choices of the parameters, a) $t_0 = 0.04$, $a_1=0.0$ and b) $t_0=0.13$, $a_1=0.33$, and the correct energy from CPIMC with the corresponding confidence interval.
 Panel (B) shows the decay of the average sign $S$ with increasing $P$ and panels (C) and (D) display the potential and kinetic 
 energy $V$ and $K$, respectively, where $E=V+K$.
 }
\end{figure}

\Table{\label{tabl1}Convergence of the energy for $N=4$, $\lambda=1.3$ and $\beta=5.0$ for selected parameter combinations shown in Fig. \ref{conv}.
}
\br

Simulation & $E$ & $V$ & $K$ & $S$\\
\mr
$^{\rm a}P=2$ & $12.1870(8)$ & $9.0157(8)$ & $3.1713(9)$ & $0.4950(5)$ \\
$^{\rm b}P=2$ & $12.3188(7)$ & $9.0840(6)$ & $3.2348(6)$  &  $0.3790(5)$  \\
$^{\rm a}P=15$  & $12.292(7)$ & $9.084(1)$ & $3.207(7)$ &   $0.02456(8)$\\
$^{\rm b}P=15$  & $12.294(4)$ & $9.0827(9)$ & $3.214(4)$ & $0.01911(4)$ \\
CPIMC & $12.293(3)$ & - & - & - & \\

\br
\end{tabular}
\item[] $^{\rm a}$ $t_0 = 0.04$, $a_1=0.0$
\item[] $^{\rm b}$ $t_0=0.13$, $a_1=0.33$
\end{indented}
\end{table}

In Fig.\ \ref{conv}, results are summarized for $N=4$ electrons with $\lambda=1.3$ and $\beta=5$, i.e., 
moderate coupling and low temperature, and panel (A) shows the convergence of the total energy as a function of the inverse number of propagators which
is proportional to the imaginary time step, $\epsilon \propto 1/P$. The red diamonds [a) $t_0 = 0.04$, $a_1=0.0$] and blue circles [b) $t_0=0.13$, $a_1=0.33$] denote two different combinations of free parameters 
and
exhibit a clearly different convergence behavior towards the exact result known from CPIMC, i.e., the black line. 
For $P=2$, the energy with parameter set a) is too low by almost one percent. With increasing $P$, $E$ increases and reaches a maximum around $P=6$, until the curves approach
the exact energy from above. For parameter set b), the energy converges monotonically from above and, even for $P=2$, the deviation from the CPIMC result is as small as $0.2\%$.
The selected energies which are listed in table \ref{tabl1} reveal that the total energy is converged for $P=15$ within the statistical uncertainty.
For the panels (C) and (D), the energy has been split into a potential ($V$) and kinetic ($K$) contribution.
For both parameter combinations, $V$ converges monotonically, although from different directions. In addition, parameter set b) gives a much better result for small $P$.
Panel (D) reveals, that the kinetic energy $K$ is responsible for the non-monotonous convergence of $E$ for parameter set a), which again delivers worse results for $P=2$, as compared
to the blue circles. 
Finally, panel (B) shows the average sign $S$ as a function of $1/P$. Both curves exhibit a similar decrease with an increasing number of propagators, as it is expected.
However, parameter set a) always allows for a better sign than b).
The reason for this behavior is the free parameter $t_0$, which controls the relative spacing between the three time slices of an imaginary time step $\epsilon$.
For $t_0=0.04$, there are a single small and two large steps. The latter allow for more blocking, since the corresponding decay length $\lambda_{t_1\epsilon}$ in the diffusion
matrices is large as well. For $t_0 = 0.13$, on the other hand, there are three nearly equal steps, each of which with a smaller decay length than the two large ones for parameter set a).
Therefore, less blocking is possible and more determinants with a negative sign appear in the Markov chain.

\begin{figure}[]
 \centering
 \includegraphics[width=0.49\textwidth]{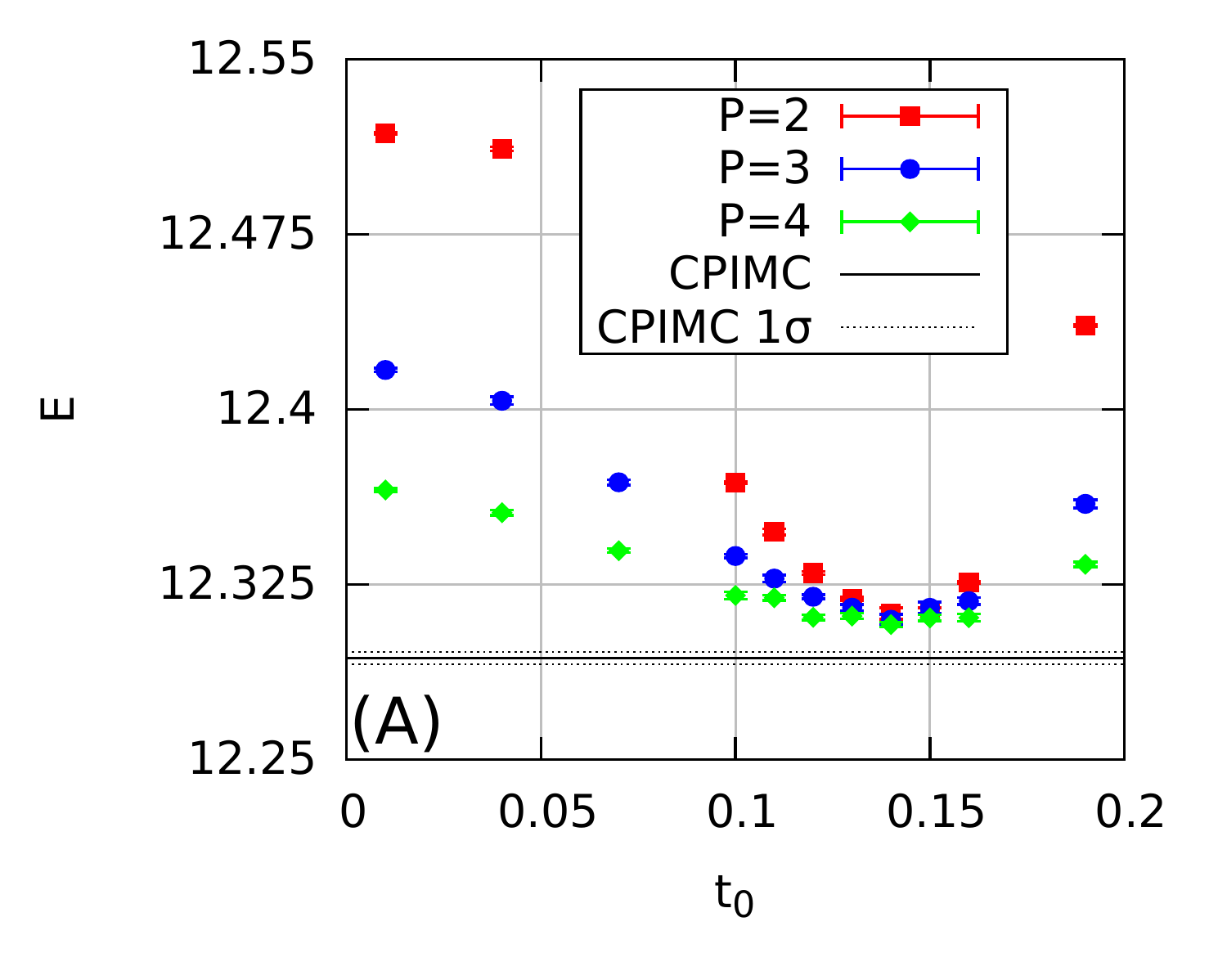}
  \includegraphics[width=0.49\textwidth]{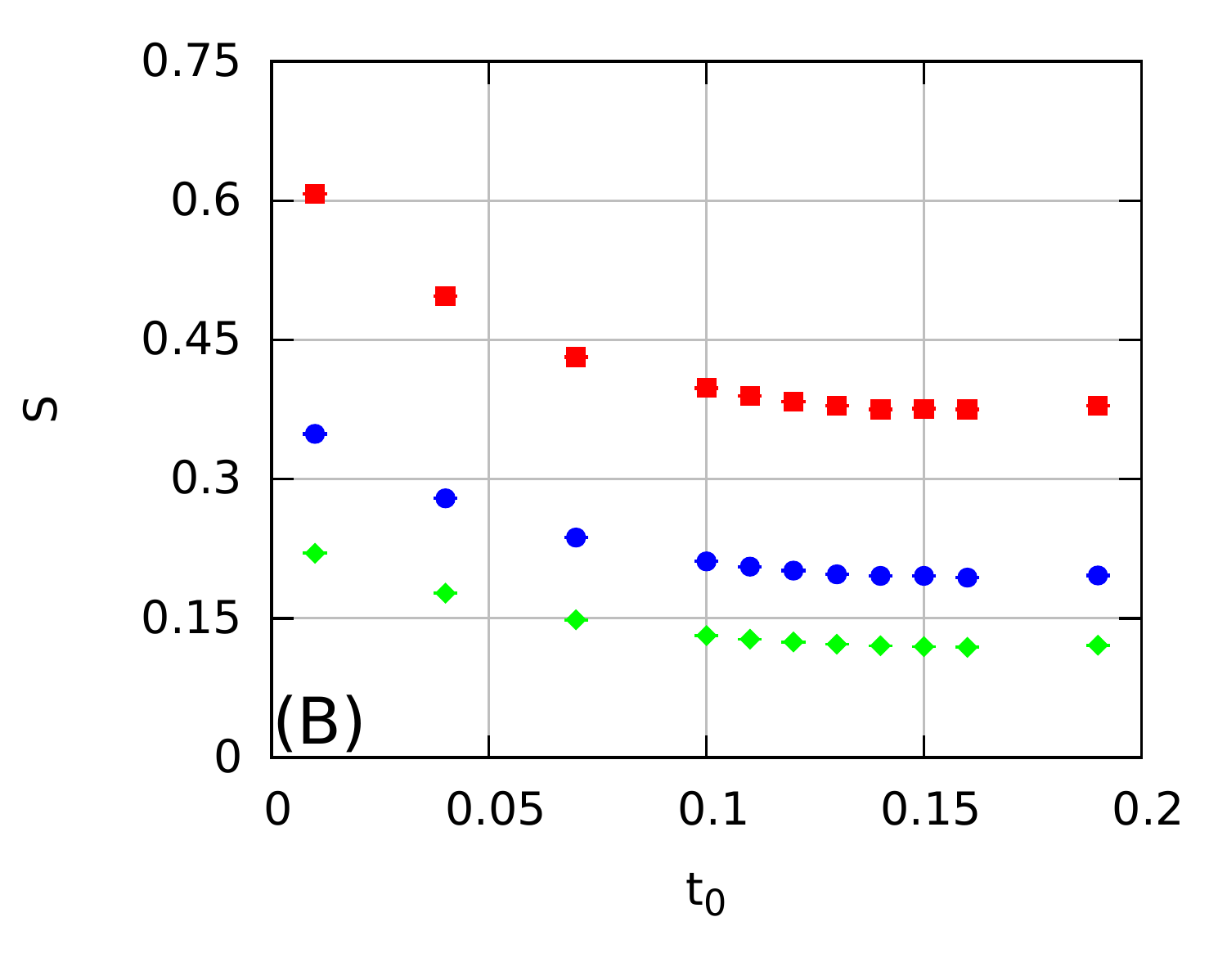}
 \caption{\label{Ps}Influence of the relative interslice spacing $t_0$ for $N=4$, $\lambda=1.3$ and $\beta=5.0$ -- 
 In the left panel, the total energy is plotted versus the free parameter $t_0$ for $P=2$, $P=3$ and $P=4$.
 The right panel shows the behavior of the average sign.
 }
\end{figure}

The different convergence behaviors of the two free parameter combinations for small $P$ leads to the question how to choose $t_0$ and $a_1$ for optimal results.
To provide an answer, we consider the same system as in Fig.\ \ref{conv}, and investigate the accuracy of the total energy as a function of $t_0$, for a fixed $a_1 = 0.33$.
The simulation results are shown in the left panel of Fig. \ref{Ps} for $P=2$ (red squares), $P=3$ (blue circles) and $P=4$ (green diamonds).
All three curves exhibit a similar decay towards the exact value starting from small $t_0$, followed by a minimum around $t_0 = 0.14$ and finally an increasing 
error for larger values. We note that as few as two propagators allow for an accuracy of $|\Delta E|/E < 2\times10^{-3}$ for the best choice of the free parameters.
Fig.\ \ref{Ps} (B) shows the dependency of the average sign $S$ on $t_0$.
Again, we observe that $S$ decreases with increasing $t_0$ as explained during the discussion of Fig.\ \ref{conv}.
In addition, it is revealed that the combination of $P=4$ and $t_0=0.01$ leads to a larger sign than $P=3$ and $t_0>0.10$.
However, the optimum free parameters allow for a higher accuracy even for $P=2$, compared to small $t_0$ with more propagators.
Therefore, it turns out to be advantagous to use the fourth order factorization with the two free parameters despite the smaller average sign for the same $P$ compared
to the factorization with only a single daughter slice for each propagator, i.e., $t_0 = 0.0$.

Finally, we mention that the optimal choice of $a_1$ and $t_0$ depends on the observable of interest.
  \begin{figure}[]
 \centering
 \includegraphics[width=0.49\textwidth]{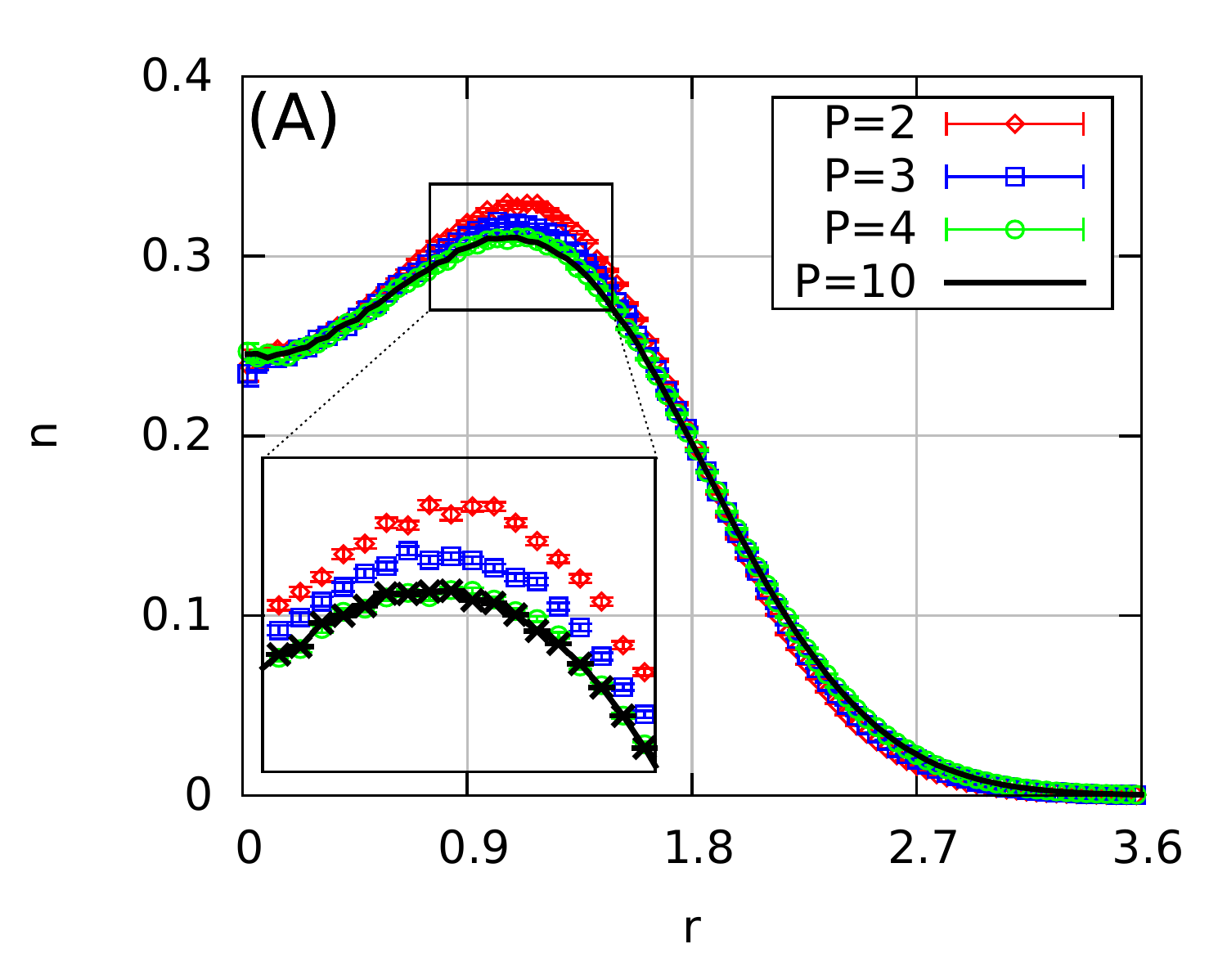}
  \includegraphics[width=0.49\textwidth]{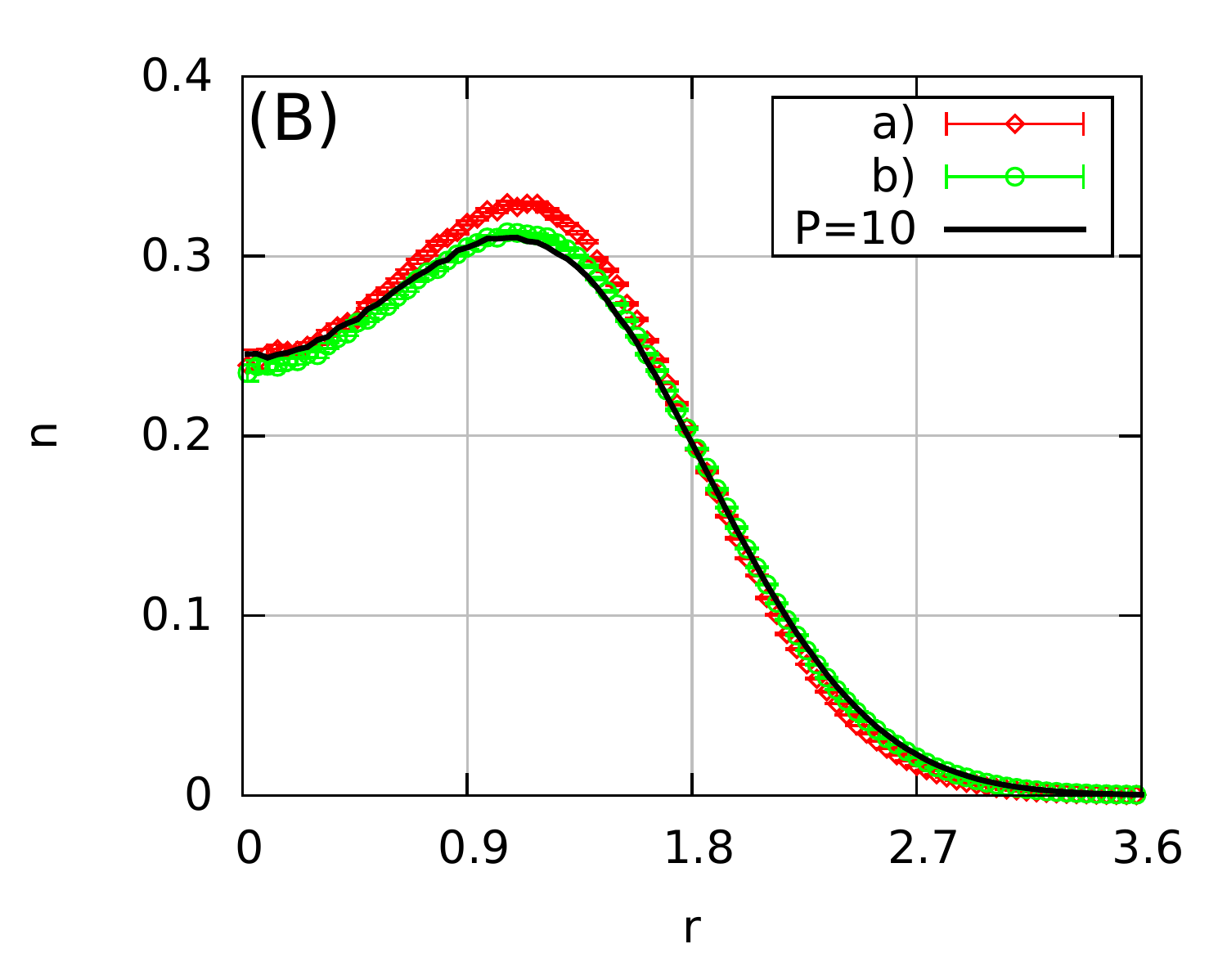}
 \caption{\label{dens}Convergence of the radial density for $N=4$, $\lambda=1.3$ and $\beta=5.0$ -- 
 The radial density $n$ is plotted versus the distance to the center of the trap, $r$.
 In panel (A), the free parameters are chosen as $t_0 = 0.13$ and $a_1 = 0.33$ and the convergence with $P$ is illustrated.
 Panel (B) compares two different sets of free parameters, a) $t_0 = 0.13$ and $a_1 = 0.33$ and b) $t_0 = 0.04$ and $a_1 = 0.0$, for $P=2$.
 }
\end{figure}
In Fig.\ \ref{dens}, we investigate the effects of the free parameters on the convergence of the radial density distribution $n(r)$ for the same system as in Figs.\ \ref{conv} and \ref{Ps}.
The left panel shows $n$ as a function of the distance to the center of the trap, $r$, for four different $P$ and the parameter combination $a_1 = 0.33$ and $t_0 = 0.13$, which
has been proven to allow for nearly optimum energy values at $P=2$, cf.\ Fig.\ \ref{Ps}.
The black curve corresponds to $P=10$ and is converged within statistical uncertainty.
For $P=2$ (red diamonds), there appear significant deviations to the latter, in particular $n$ is too large around the maximum $r\approx1.25$ and too small at the boundary of the system.
The $P=3$ results (blue squares) exhibit the same trends although the differences towards the black curve are reduced.
Finally, the density for $P=4$ (green circles) cannot be distinguished from the converged data within the error bars.
The right panel compares the density for $P=2$ with two different combinations of free parameters.
The red diamonds [parameter set a)] correspond to the curve from the left panel and the green circles [parameter set b)] to $a_1=0.0$ and $t_0 = 0.04$.
The latter parameters clearly allow for a density distribution which is much closer to the exact results than the $a_1$ and $t_0$ values which provide the optimal energy.

\subsection{Temperature dependence}
In the last section, we have demonstrated that the optimal choice of the free parameters $a_1$ and $t_0$ allows for the calculation of energies with an accuracy of $0.1\%$
with as few as two propagators, even at a relatively low temperature, $\beta=5.0$. However, with decreasing $T$ (i.e., increasing $\beta$) the number of required propagators must be increased
to keep the commutator error fixed.
  \begin{figure}[]
 \centering
 \includegraphics[width=0.49\textwidth]{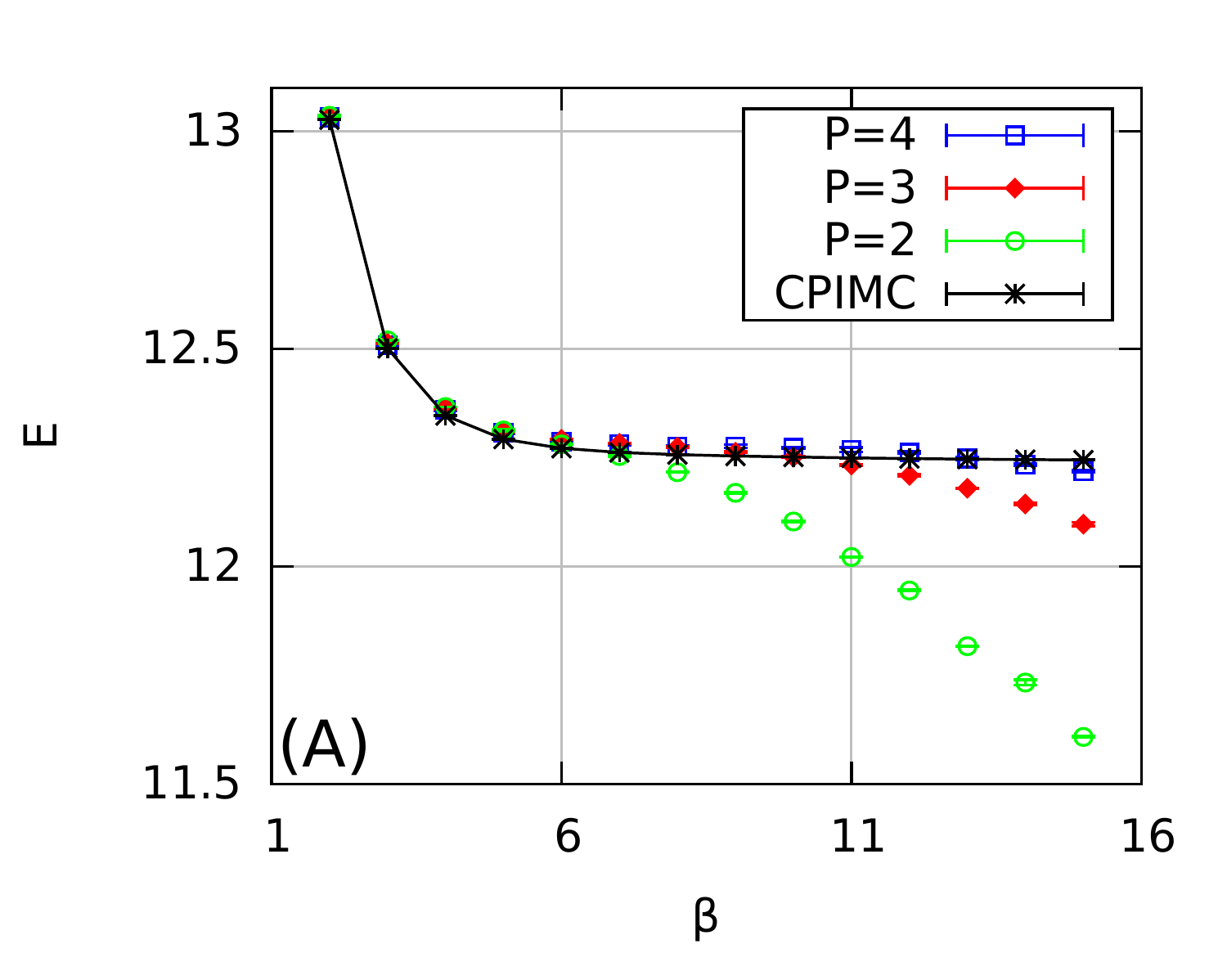}
  \includegraphics[width=0.49\textwidth]{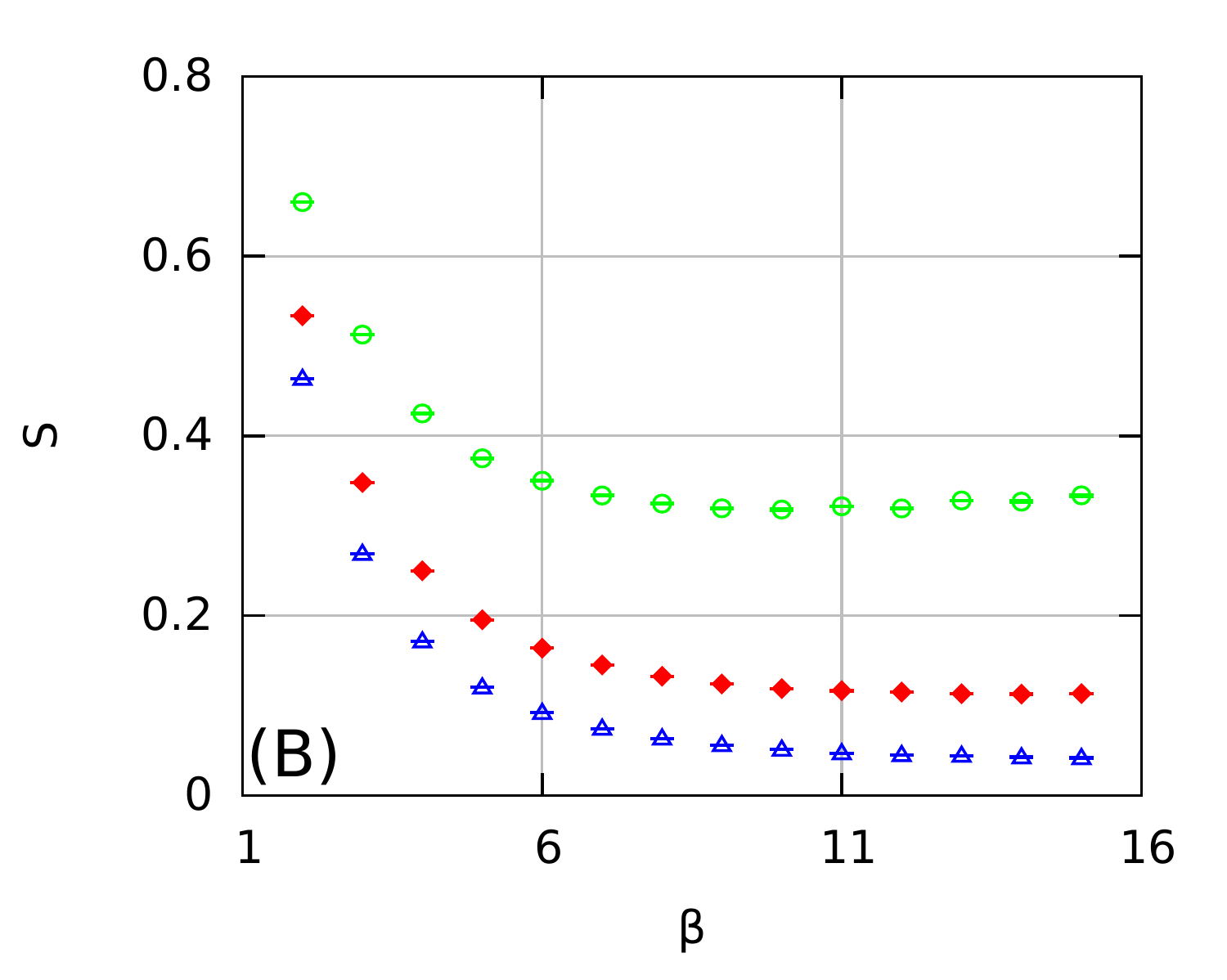}
 \caption{\label{beta}Temperature dependence for $N=4$ and $\lambda=1.3$ with $t_0= 0.14$ and $a_1=0.33$ -- 
 In the left panel, the total energy is plotted versus the inverse temperature $\beta$ for $P=2$, $P=3$ and $P=4$ propagators and 
 compared to exact CPIMC results.
 The right panel shows the behavior of the sign.
 }
\end{figure}
In Fig.\ \ref{beta}, we investigate the effect of a decreasing temperature on the accuracy provided by a few propagators $P$ for $N=4$ electrons at indermediate coupling, $\lambda=1.3$.
The left panel shows the total energy $E$ as a function of the inverse temperature $\beta$.
We compare results for $P=2$ (green circles), $P=3$ (red diamonds) and $P=4$ (blue triangles) to exact results from CPIMC (black stars).
At larger temperature, $\beta \le 7.0$, all four datasets nearly coincide and exhibit the expected decrease towards the energy of the ground state.
With increasing $\beta$, the $P=2$ results exhibit an unphysical drop because two propagators are not sufficient and the commutator errors become more significant.
The red and blue curves exhibit a qualitatively similar trend, however, the energy drop is weaker and shifted to lower temperature.
Even at $\beta=10.0$, which is already very close to the ground state, 
three propagators allow for an accurate description of the system.

In the right panel of Fig.\ \ref{beta}, the average sign $S$ is plotted versus the inverse temperature.
At small $\beta$, the wavefunctions of the electrons do not overlap and, hence, the system is not degenerate. With decreasing temperature,
exchange effects become increasingly important which leads to a decrease of $S$.
However, while for standard PIMC the sign is expected to exponentially decrease with $\beta$, $S$ seems to converge for PB-PIMC with $P=3$ and $P=4$ and exhibits an even slightly
non-monotonous behavior for $P=2$.
The application of antisymmetric propagators leads to a competition with respect to $S$ and $\beta$. On the one hand, with increasing inverse temperature off-diagonal matrix elements
are increased, which leads to more negative determinants and, therefore, more negative weights in the Markov chain.
On the other hand, the thermal wavelengths $\lambda_{t_1\epsilon}$ and $\lambda_{2t_0\epsilon}$ are increasing with $\beta$, which makes the blocking of large diagonal and off-diagonal
elements more effective.
Hence, the sign can even become larger with $\beta$ once the system has reached the ground state, because the particle distribution remains constant while more elements in the diffusion
matrix compensate each other in the determinants.

We conclude that few propagators allow for the calculation of accurate results up to low temperature, $\beta \le 10.0$. For higher $\beta$, the system is in its
ground state and finite temperature path integral Monte Carlo is no longer the method of choice.

\subsection{Dependence on the coupling strength}
In the previous sections, we have restricted ourselves to the investigation of small systems to illustrate the convergence and sign behavior depending on relevant parameters.
In this section, we demonstrate that PB-PIMC allows for the calculation of accurate results at parameters where no other ab initio results have been reported, so far.
  \begin{figure}[]
 \centering
 \includegraphics[width=0.47\textwidth]{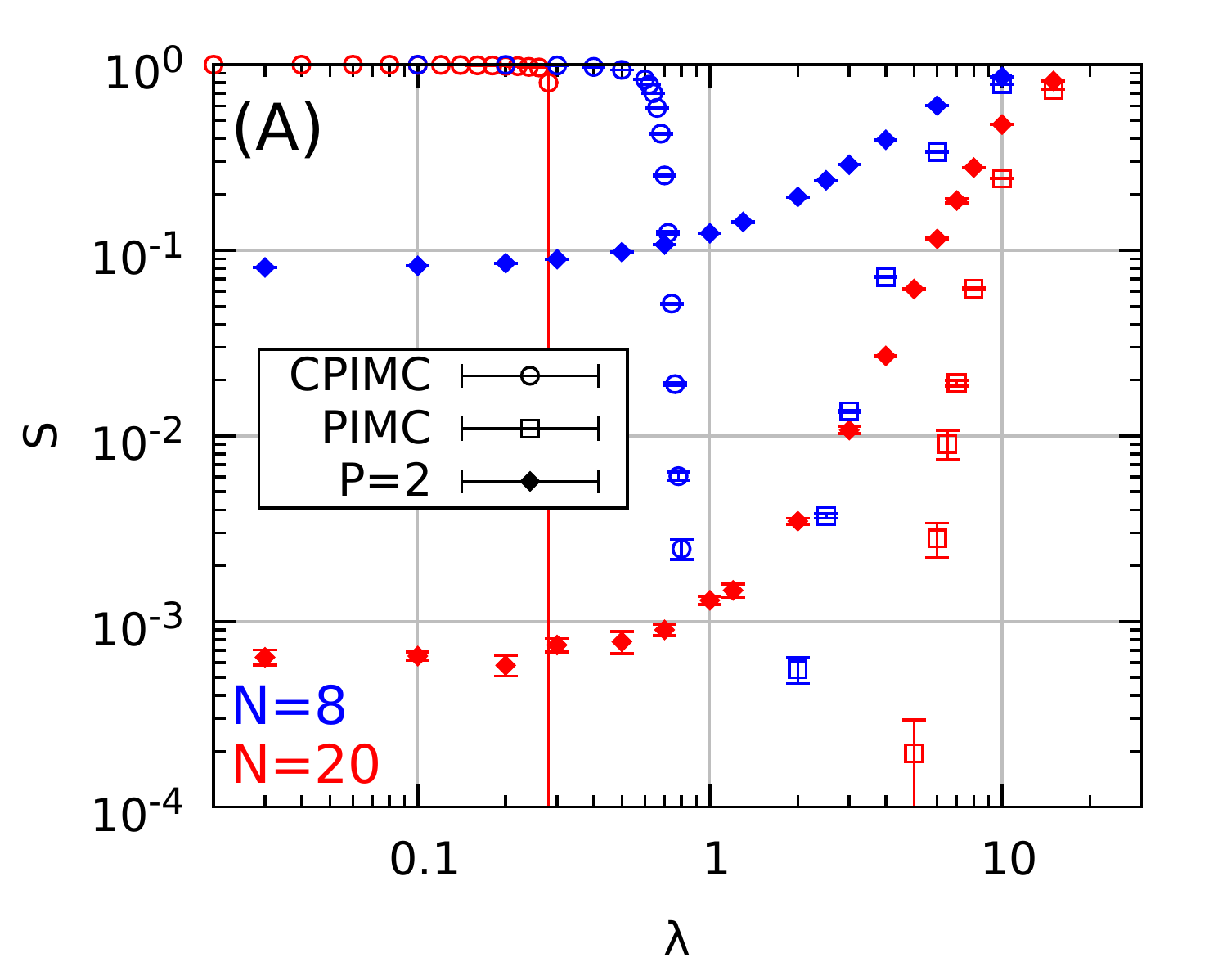}
  \includegraphics[width=0.49\textwidth]{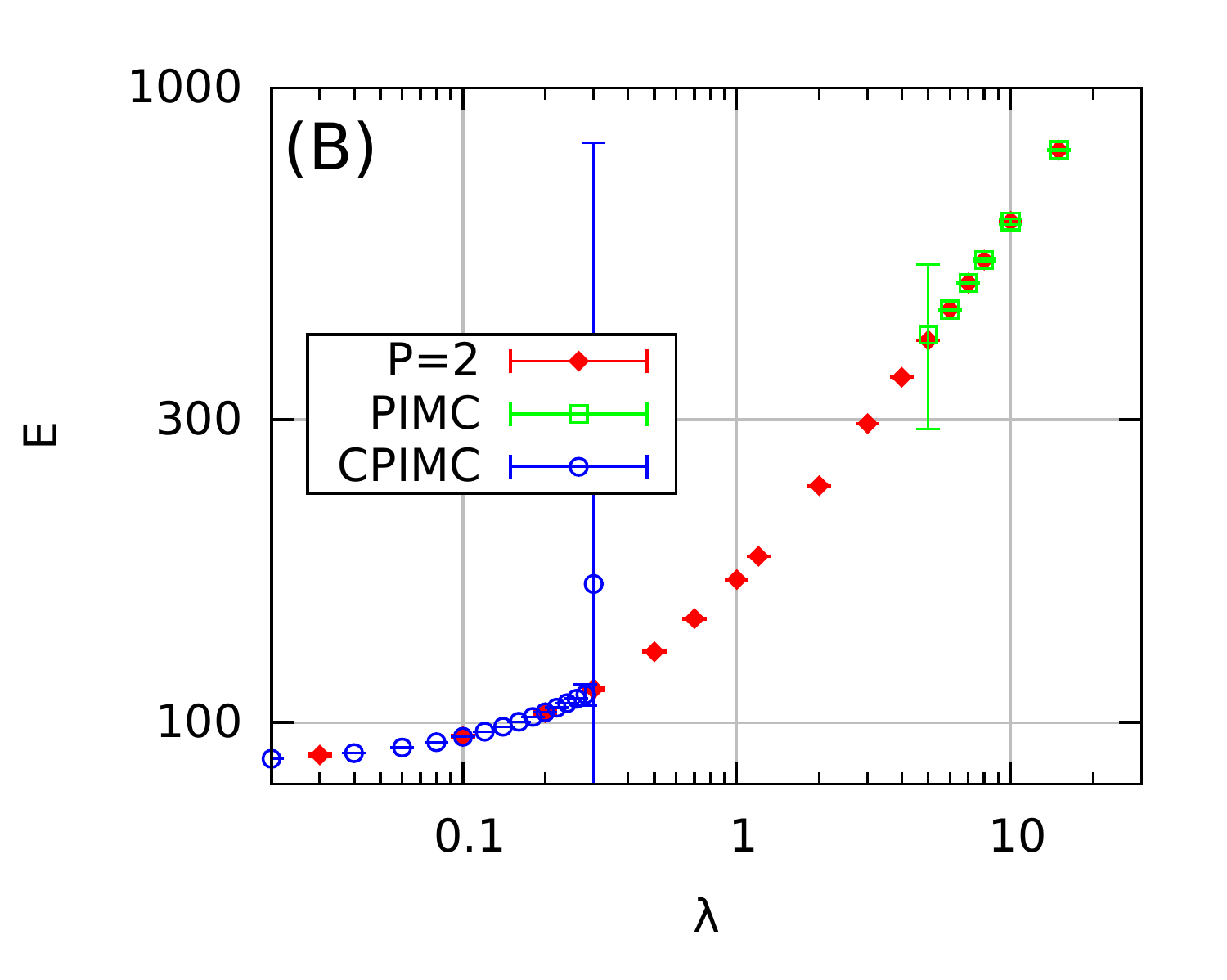}\\
  \includegraphics[width=0.49\textwidth]{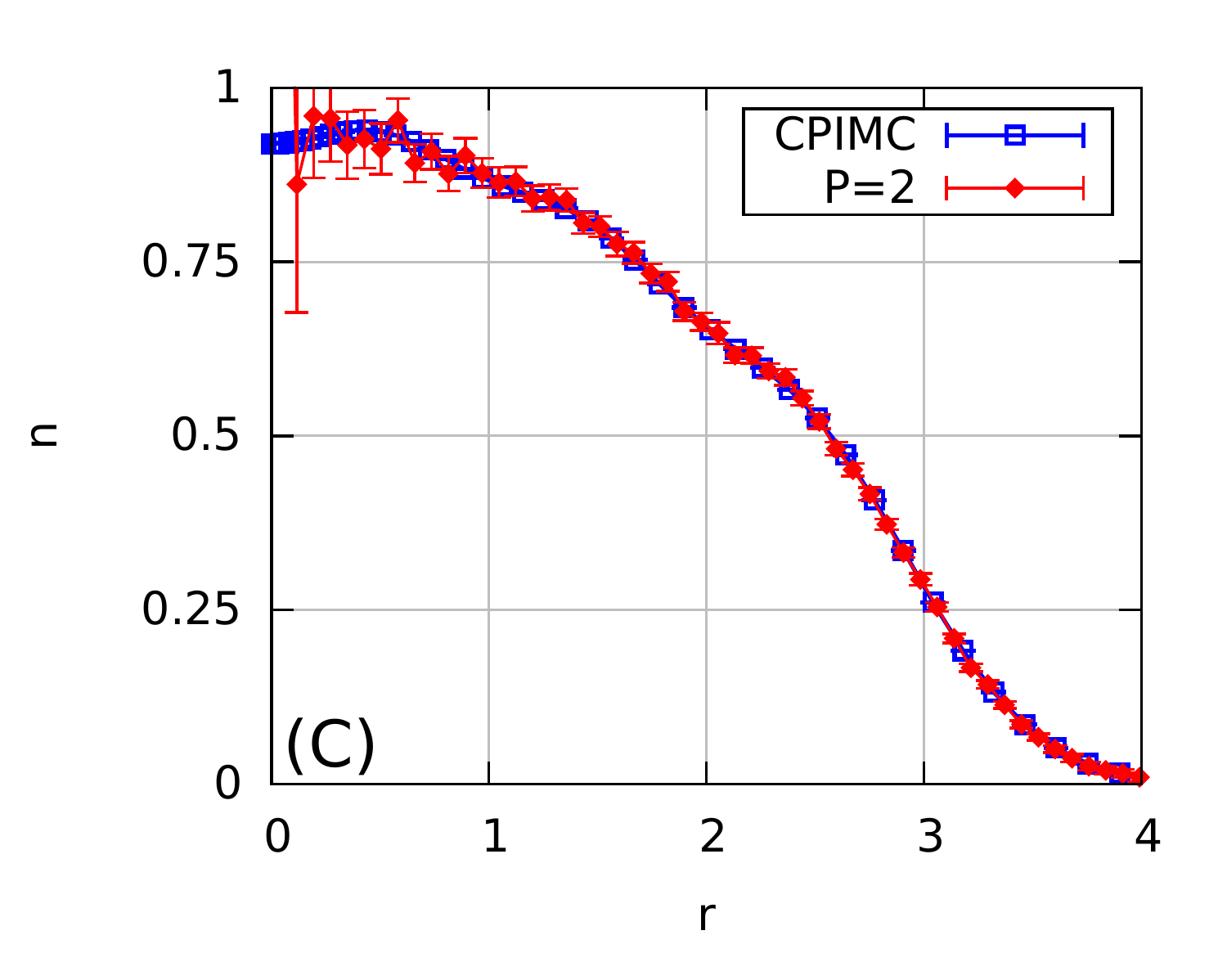}
  \includegraphics[width=0.49\textwidth]{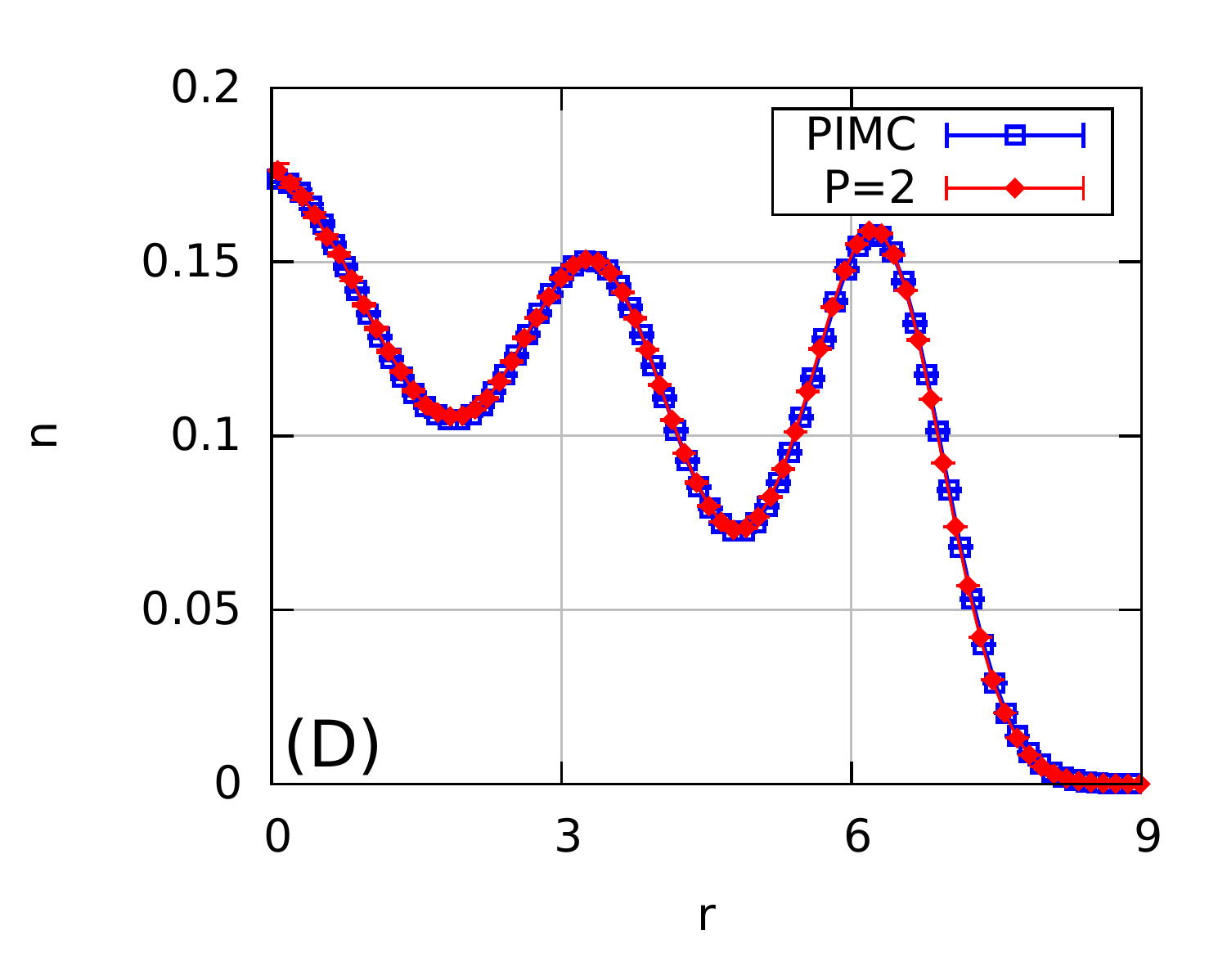}
 \caption{\label{coupling}Coupling dependence for $N=8$ and $N=20$ at $\beta=3.0$ -- 
 Panel (A) shows the average sign as a function of $\lambda$ for CPIMC, PIMC and PB-PIMC with $N=8$ (blue symbols, parameter set $t_0=0.14$ and $a_1=0.33$) 
 and $N=20$ (red symbols, parameter set $t_0=0.10$ and $a_1=0.33$) and panel (B) the corresponding total energies, $E$, for the latter.
 In panels (C) and (D), the radial density $n$ is plotted versus the distance to the center of the trap, $r$, for $N=20$ with $\lambda=0.1$ and $\lambda=15.0$, respectively, and the parameter set $a_1=0.0$ and $t_0=0.04$.
 }
\end{figure}
Fig.\ \ref{coupling} shows results for $N=8$ and $N=20$ electrons at $\beta=3.0$ over a wide range of coupling parameters, $\lambda$.
In panel (A), the average sign $S$ is plotted versus $\lambda$ for standard PIMC (squares), CPIMC (circles) and the present PB-PIMC (diamonds) with $P=2$ and the parameter sets $t_0=0.14$ and $a_1=0.33$ ($N=8$, blue symbols) and $t_0=0.10$ and $a_1=0.33$ ($N=20$, red symbols),
which are known to allow for accurate energies, cf.\ Fig.\ \ref{Ps}.
It is well understood that PIMC allows for the simulation of strongly coupled fermions, where exchange effects do not play a dominant role.
With decreasing $\lambda$, the sign exhibits a sharp drop and the sign problem prevents the simulation within feasible computation time for $\lambda\le2.0$ and $\lambda\le5.0$, respectively.
Evidently, larger systems lead to a more severe decrease of $S$ at larger coupling strength.
CPIMC, on the other hand, can be interpreted as a Monte Carlo simulation on a perturbation expansion around the ideal quantum system, i.e., $\lambda=0.0$.
Hence, the method efficiently provides exact results for small coupling, where the system is close to an ideal one.
For $N=20$ around $\lambda \approx 0.3$, the sign almost instantly drops from $S\approx0.97$ towards zero, and CPIMC is no longer applicable, without further approximation.
This means that, in particular for larger systems, there have only been results for systems that are a) almost ideal or b) so strongly coupled that fermions and bosons lead to nearly equal physical properties.
The physically particularly interesting regime where Coulomb correlations and Fermi statistics are significant simultaneously, has remained out of reach.

However, the average sign from PB-PIMC exhibits a much less severe drop with decreasing $\lambda$ than standard PIMC and saturates for $\lambda\le0.7$.
For $N=8$, the average sign remains above $S=0.08$, which allows for good accuracy with relatively low effort.
The small sign, $S\sim10^{-3}$, for $N=20$ indicates that the simulations are computationally involved but, in contrast to PIMC and CPIMC, still feasible.
In panel (B) of Fig.\ \ref{coupling}, the total energy $E$ for $N=20$ is plotted versus $\lambda$ over the entire coupling range and the
statistical uncertainty from the PB-PIMC results is smaller than the size of the data points.
Both, at small and large $\lambda$, the $P=2$ results are in excellent agreement with the exact energy known from the other methods and, in addition,
results are obtained for the particularly interesting transition region (region [II] in Fig.\ \ref{intro}).
In panel (C), we show the radial density for $N=20$ and low coupling, $\lambda=0.10$, calculated with the parameter set $t_0=0.04$ and $a_1=0.0$, which has been proven effective for accurate densities $n(r)$.
The PB-PIMC results (red diamonds) are in excellent agreement with the exact CPIMC data (blue squares) over the entire $r$-range.
For completeness, we mention that this combination of parameters allows for an approximately three times as high sign as the choice from panels (A) and (B), which was choosen
to allow for a good energy.
Panel (D) shows the density of a strongly coupled system, $\lambda=15.0$, and $N=20$.
Again, the two propagators already provide very good agreement with the exact curve.
In Fig.\ \ref{intro} (B), we have shown density profiles for coupling parameters over the entire coupling range.
At $\lambda=15$ (red pluses), there are three distinct shells and the physical behavior is dominated by the strong Coulomb repulsion.
Decreasing the coupling to $\lambda=5$ (green bars) leads to a reduced extension of the system, and the three shells exhibit a much larger overlap.
At indermediate coupling, $\lambda=2$ (blue crosses), both the interaction and fermionic exchange govern the system. The density profile is still
significantly more extended than the ideal pendant, but $n$ exhibits modulations instead of a flat curve.
Decreasing the repulsion further to $\lambda=0.7$ (pink circles) leads to a further reduction of the extension. However, $n$ does
not approach a Gaussian-like profile as for ideal boltzmannons or bosons, but continues to exhibit the density modulations which are characteristic for fermions.
For $\lambda=0.1$, the system is almost ideal and the density is completely dominated by the quantum statistics.
  \begin{figure}[]
 \centering
 \includegraphics[width=0.49\textwidth]{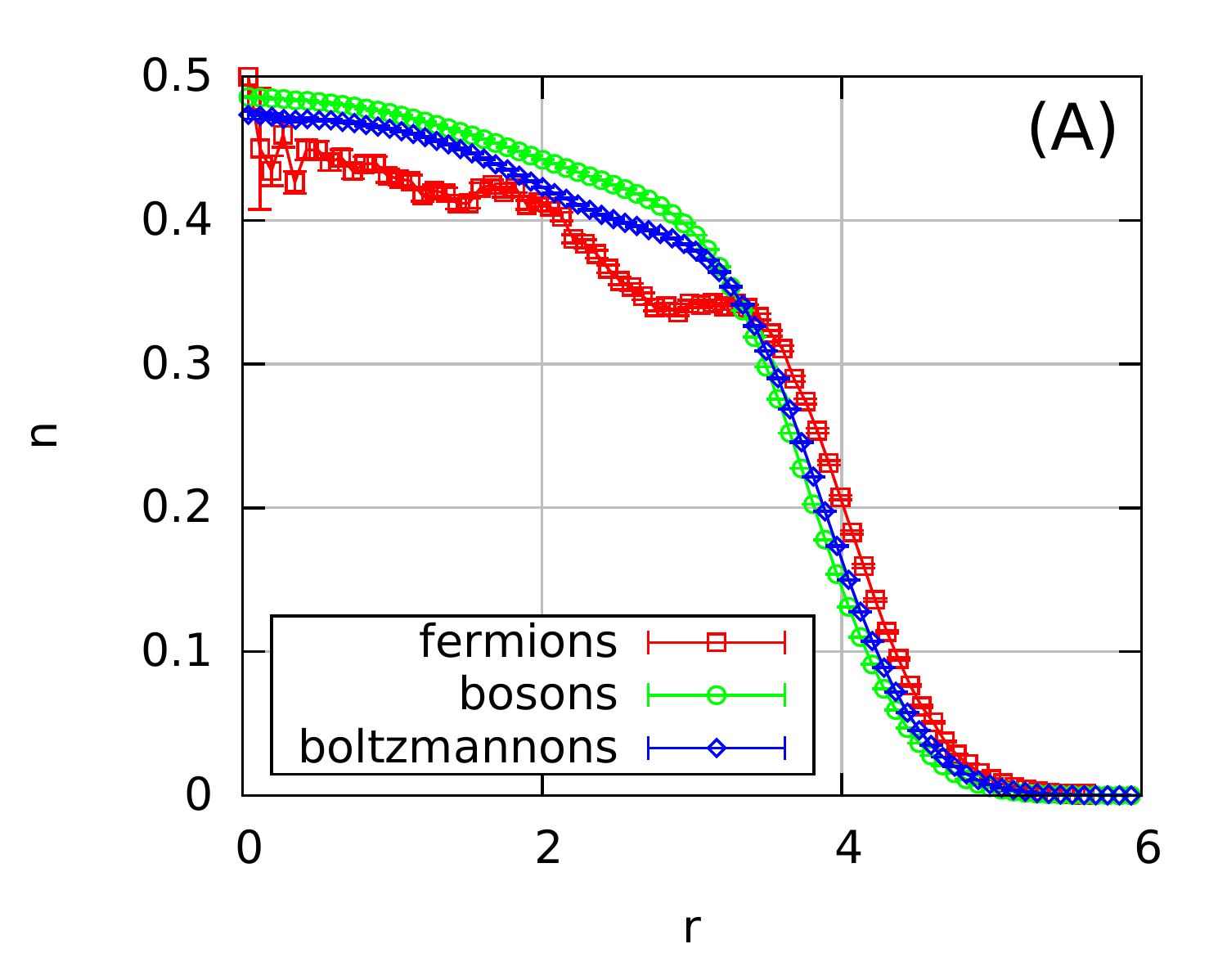}
  \includegraphics[width=0.49\textwidth]{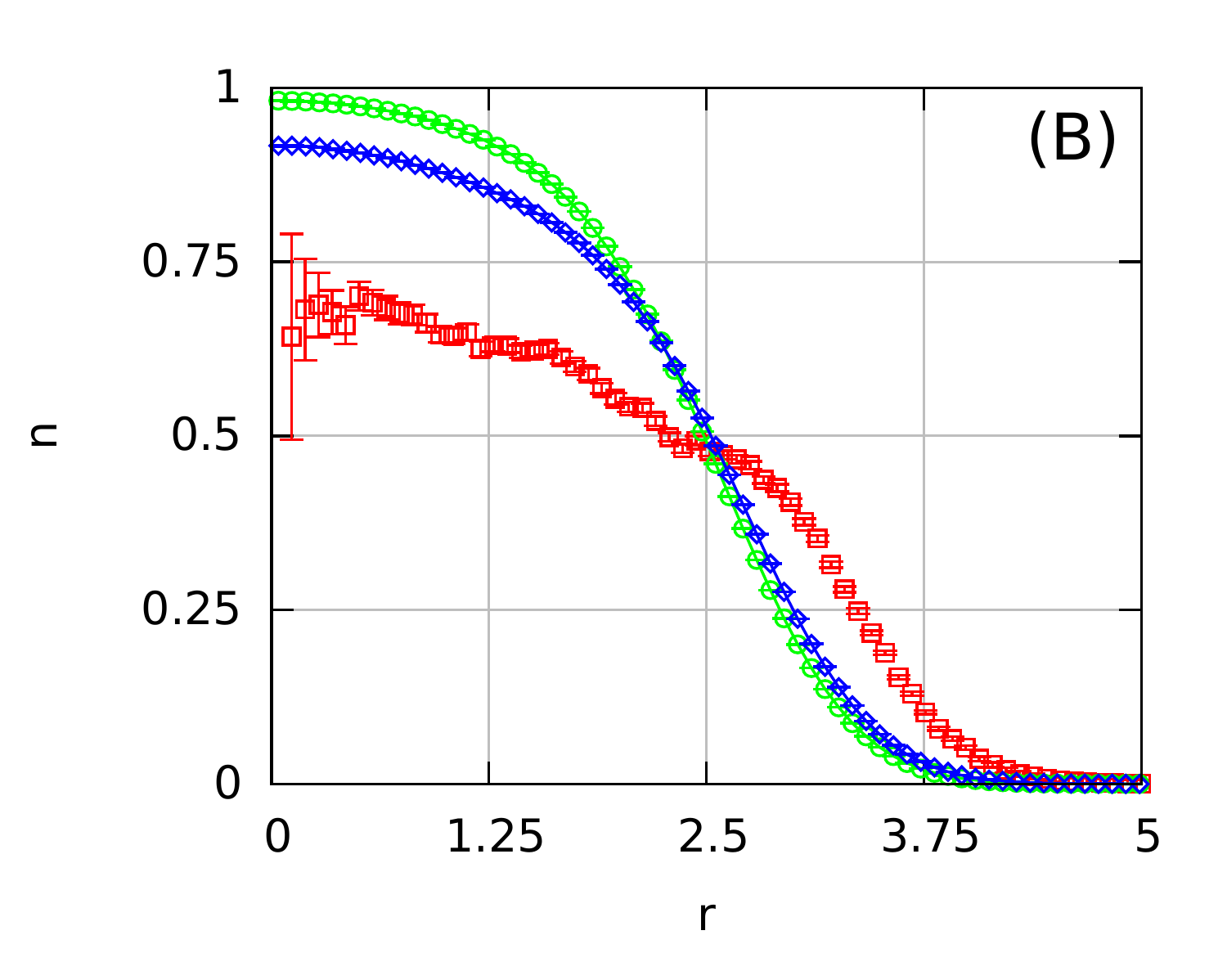}
 \caption{\label{coupling2}Influence of quantum statistics for $N=20$ and $\beta=3.0$ -- 
We show the radial density $n(r)$ for Fermi-, Bose- and Boltzmann-statistics in the transition region for $\lambda=2.0$ (A) and $\lambda=0.7$ (B).
 }
 \end{figure}

Finally, in Fig.\ \ref{coupling2} we compare density profiles for $N=20$ particles at $\beta=3.0$ with Fermi-, Bose- and Boltzmann statistics.
Panel (A) shows results for intermediate coupling, $\lambda=2.0$.
The distinguishable boltzmannons (blue diamonds) exhibit a nearly flat profile without any shell structure, i.e., a liquid-like behavior.
The bosonic particles (green circles) lead to an even smoother curve, with a slightly reduced extension of the system.
For fermions (red squares), on the other hand, the exchange already plays a significant role, as the particles 
exhibit an additional repulsion due to the Pauli principle,
and $n$ decays only at larger $r$. In addition, the fermionic density profile exhibits distinct modulations.
In panel (B), we show a comparison for smaller coupling, $\lambda=0.7$. 
Again, the boltzmannons and bosons lead to smooth density profiles which are very similar, despite a reduced extension of the Bose-system and an
increased density around the center of the trap. The fermions exhibit a different behavior as the system is significantly more
extended and the density profile again features distinct modulations.

In conclusion, we have presented ab initio results for the energy and the density for up to $20$ electrons over the entire coupling range. 
A comparison with standard PIMC and CPIMC has revealed excellent agreement in both the limits of weak and strong coupling.
A more detailed investigation of the transition from the classical to the degenerate regime, including systematic comparisons with bosons and boltzmannons, 
is beyond the scope of this work and will
be published elsewhere.

\subsection{Particle number dependence}
In the last section, we have shown that the sign problem is more severe for larger systems, cf. Fig.\ \ref{coupling} (A).
Here, we provide a more detailed investigation of the performance of our method in dependence on the particle number.
  \begin{figure}[]
 \centering
 \includegraphics[width=0.49\textwidth]{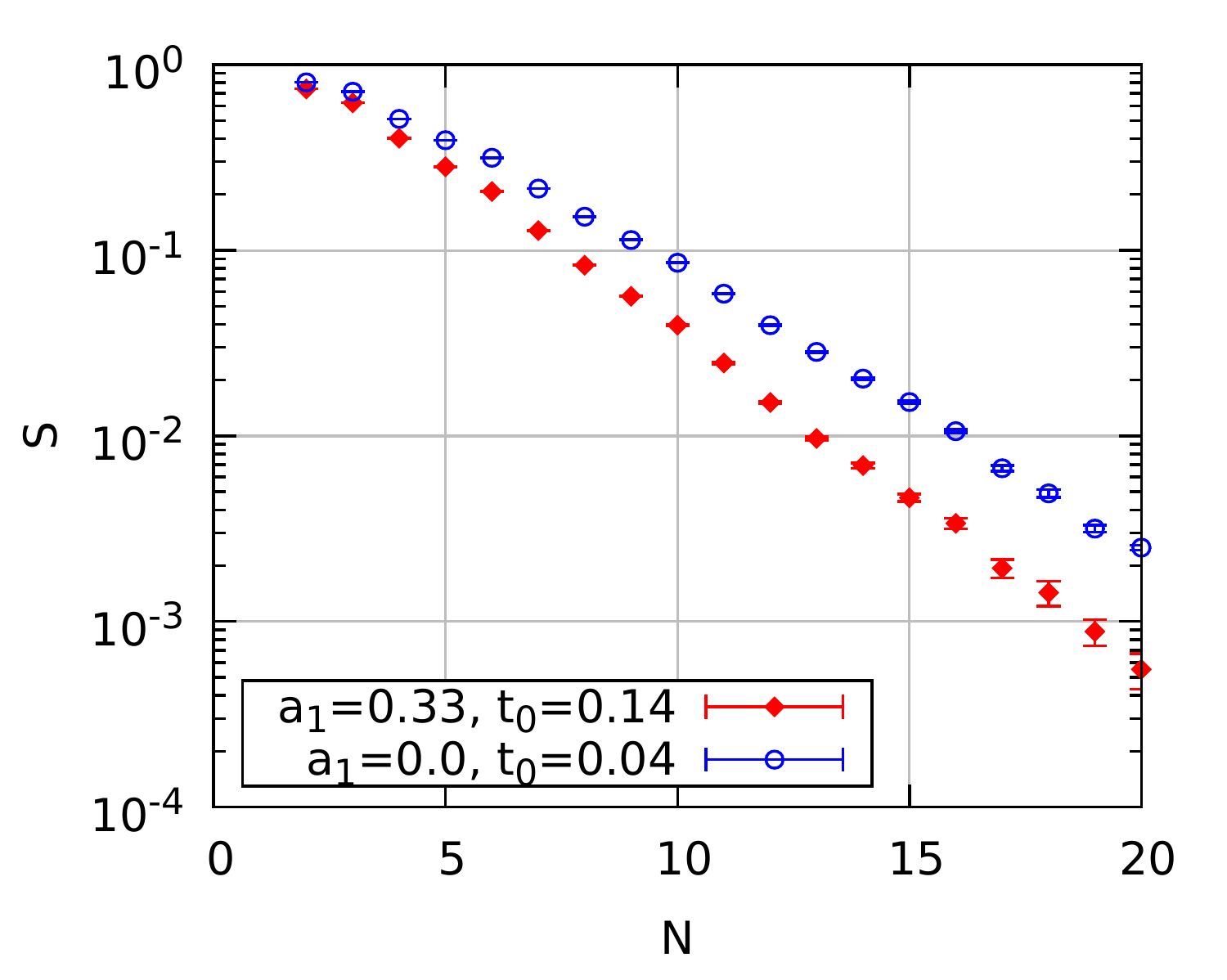}
 \caption{\label{N}Particle number dependence of the average sign for $\lambda=0.1$ and $\beta=3.0$  
 and two different 
 combinations of the simulation parameters.
 }
\end{figure}
In Fig.\ \ref{N}, the average sign $S$ is plotted versus $N$ for $\lambda=0.1$ and $\beta=3.0$, i.e., a very degenerate system, with two different combinations of free parameters.
It is revealed that $S$ exhibits an exponential decay with the system size and, as usual, the smaller $t_0$ leads to a more effective blocking.
Therefore, the PB-PIMC approach still suffers from the fermion sign problem, 
and feasible system sizes for $2D$ quantum dots at weak coupling are limited to $N\le30$.
This is a remarkable result since standard PIMC simulations for $\lambda=0.1$ and $\beta=3.0$ are possible only for $N\le4$.

\section{Discussion}
In summary, we have presented a novel approach to the path integral Monte Carlo simulation of degenerate fermions at finite temperature by combining a fourth-order factorization of the 
density matrix with a full antisymmetrization between all imaginary time slices. 
The latter allows to merge $3PN!$ configurations from the standard PIMC formulation into a single configuration weight, thereby efficiently grouping together permutations 
of opposite signs which leads to a significant relieve of the fermion sign problem.
To efficiently run through the resulting configuration space at arbitrary system parameters, we have modified the widely used continuous space worm algorithm by introducing an extended ensemble with
open configurations and by temporarily constructing artificial trajectories.
We have demonstrated the capabilities of our method by simulating up to $N=20$ electrons in a quantum dot.
It has been revealed that the (empirical) optimal choice of the free parameters $a_1$ and $t_0$ from the fourth order factorization allows for the
calculation of energies with an accuracy of $0.1\%$ even for just two
propagators. For completeness, we mention that different observables lead to different optimal parameters.
We have concluded, that it appears to be favourable to use two instead of a single daughter time slice for each time step $\epsilon$, despite the reduced sign for the same number of propagators.

The investigation of the temperature dependence of the convergence with respect to the number of time steps $P$ has revealed, that as few as three propagators
are sufficient to accurately simulate fermions, up to $\beta \le 10.0$. For larger inverse temperatures, the system approaches its ground state and
finite temperature path integral Monte Carlo techniques are no longer the methods of choice.

To demonstrate that our PB-PIMC approach allows for the calculation of accurate results for systems beyond the capability of any other quantum Monte Carlo technique,
we have simulated $N=20$ electrons at relatively low temperature, $\beta=3.0$, and arbitrary coupling strength. CPIMC excells at weak coupling and provides exact results for $\lambda < 0.3$, i.e.,
in the region where the systems are still close to the ideal case.
Standard PIMC, on the other hand, is applicable at strong coupling $\lambda \ge 5.0$ where exchange effects 
are not yet dominating, until the rapid decrease of the sign renders any simulation unfeasible.
For PB-PIMC, the sign converges for $\lambda \le 0.7$ and, hence, computations are possible at arbitrary degeneracy, in particular, in the physically most interesting
transition region between classical and ideal quantum behavior.
We find excellent agreement with both PIMC and CPIMC in both the limits of strong and weak coupling.
Finally, we have demonstrated that PB-PIMC still suffers from the fermion sign problem, since, as expected, $S$ exponentially decreases with the particle number.

A possible future application of PB-PIMC to the quantum dot system might include the investigation of the transition from the classical to the degenerate quantum regime, in particular a systematic
comparison of fermions to bosons and boltzmannons. Furthermore, it could be interesting to extend the considerations to $3D$ confinements, e.g.\ \cite{pludwig,dornheim},
and study the impact of quantum statistics on structural transitions \cite{thomsen}.
In addition, we expect our method to be of interest for the future investigation of numerous Fermi systems,
including the finite temperature homogeneous electron gas \cite{brown,prl,vfil4}, two-component plasmas \cite{bonitz,morales,proton} and fermionic bilayer systems \cite{fbilayer,ludwig,fbilayer2}.

%

%

\section*{Acknowledgements}
We acknowledge stimulating discussions with T. Schoof (Kiel) and V.S. Filinov (Moscow). This work is supported by the Deutsche Forschungsgemeinschaft via SFB TR-24
project A9 and via project BO 1366/10 as well as by grant
SHP006 for CPU time at the Norddeutscher Verbund f\"ur
Hoch- und H\"ochstleistungsrechnen (HLRN).

\appendix
\section{Monte Carlo updates}\label{app}
In this appendix, we present an ergodic set of Monte Carlo updates which are based on the usual
continuous space worm algorithm \cite{bon,bon2} from standard PIMC.

  \begin{figure}[]
 \centering
 \includegraphics[width=0.49\textwidth]{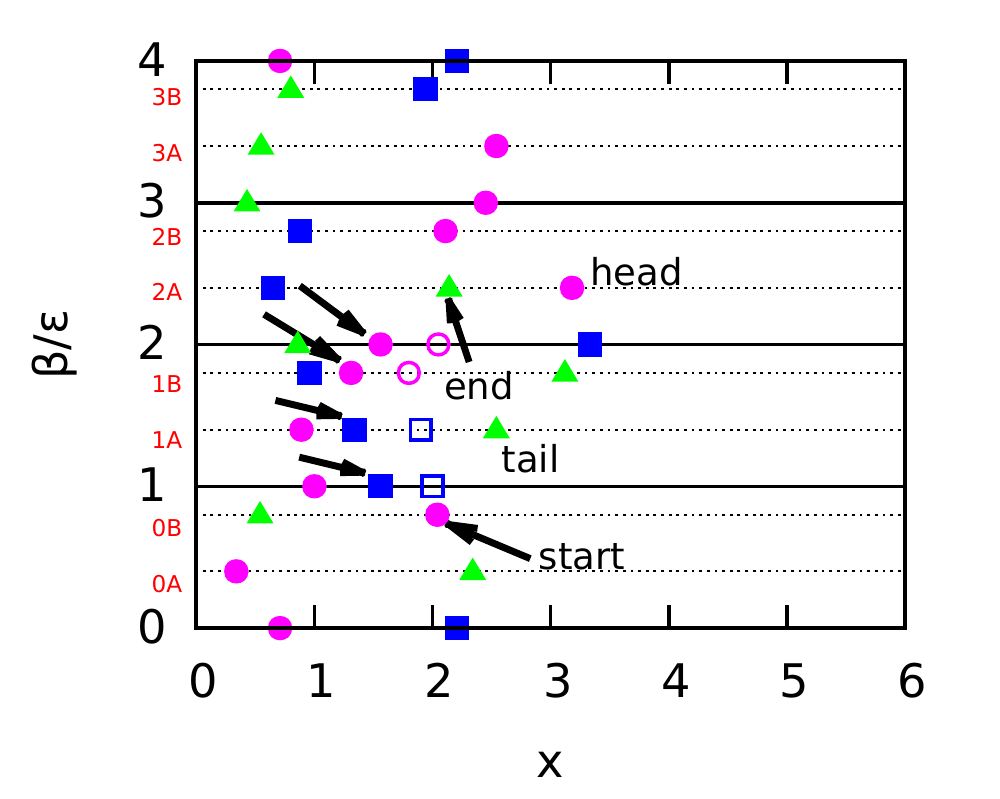}
  \includegraphics[width=0.49\textwidth]{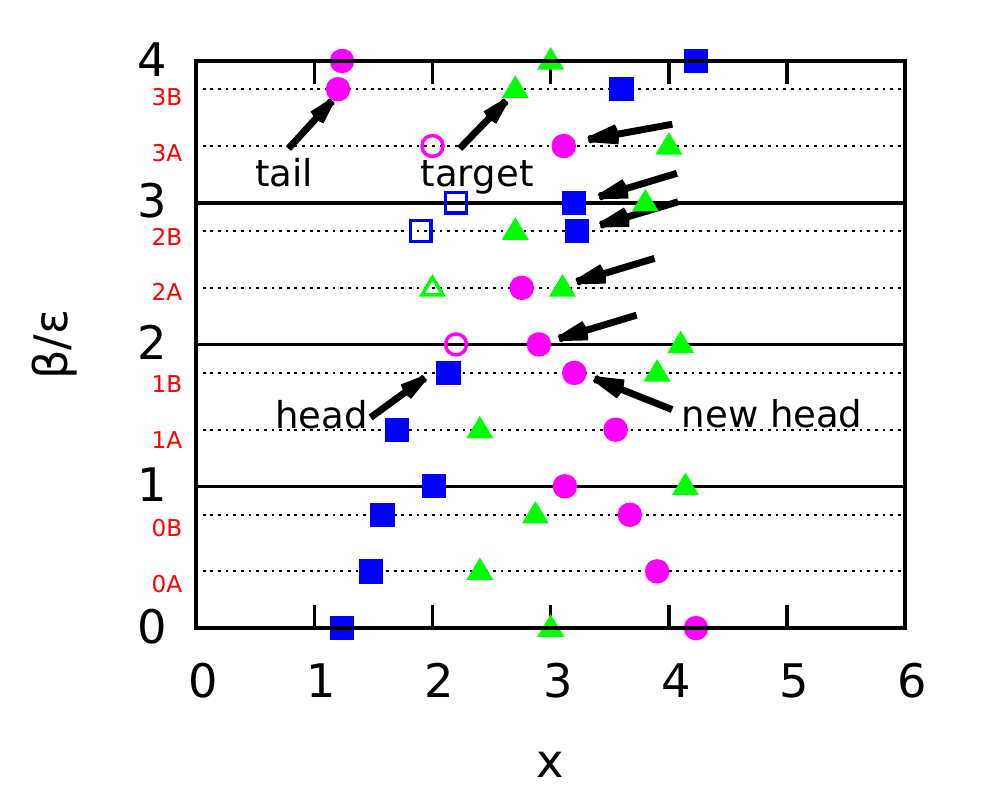}
 \caption{\label{deform}Illustration of the updates \textit{Deform} (left) and \textit{Swap} (right) --
 In the left panel, the \textit{Deform} update is executed in an open configuration. The random construction of an artificial trajectory (the beads marked by black arrows)
 is followed by the re-sampling of all beads between its first (start) and last (end) bead.
 In the right panel, the \textit{Swap} move is demonstrated. The current head is 'connected' to a random target bead on the time slice of the tail.
 }
\end{figure}

\begin{enumerate}
 \item \textbf{Deform}: This update is similar to standard PIMC techniques, e.g.\ \cite{charge}, and deforms a randomly constructed artificial trajectory.
 \begin{itemize}
  \item Select a start time $\tau_s$ uniformly from all $3P$ slices.
  \item Select a 'start' bead on $\tau_s$.
  \item Select the number of beads to be changed, $m\in[1,\tilde M]$.
  \item Select $m+1$ beads on the next slices according to 
  \begin{eqnarray}
  T_\textnormal{select} = \prod_{i=0}^m \frac{\rho_0(\mathbf{r}_i^\textnormal{old},\mathbf{r}_{i+1}^\textnormal{old},\epsilon_i)}{\Sigma_i^\textnormal{old}}  \quad ,
  \end{eqnarray}
  with $\Sigma_i^\textnormal{old}$ being the normalization and the label 'old' indicates the configuration before the update.
  \item Resample $m$ beads in the middle according to Eq.\ (\ref{connect}):
  \begin{eqnarray}
   T_\textnormal{resample} = \frac{ \prod_{i=0}^m \rho_0(\mathbf{r}_i^\textnormal{new},\mathbf{r}_{i+1}^\textnormal{new},\epsilon_i) }{ \rho_0(\mathbf{r}_0, \mathbf{r}_{m+1}, \epsilon_\textnormal{tot}) } \quad ,
  \end{eqnarray}
and $\epsilon_\textnormal{tot}$ denotes the imaginary time difference between the fixed endpoints.
 \end{itemize}
The constant $\tilde M$ is a free parameter and can be optimized to enhance the performance.
The update is self-balanced and the Metropolis solution for the acceptance probability is given by
\begin{eqnarray}
 A_\textnormal{Deform}(\mathbf{X}\to \mathbf{\tilde X}) = \textnormal{min} \left( 1, e^{-\epsilon\Delta\Phi} \prod_{i=0}^m \left| \frac{\Sigma_i^\textnormal{old}}{\Sigma_i^\textnormal{new}} \frac{ \textnormal{det} \rho_i^\textnormal{new}}{\textnormal{det}\rho_i^\textnormal{old}} \right| \right) \quad ,
\end{eqnarray}
with $\Phi$ containing both the change in the potential energy and all forces. \textit{Deform} is illustrated in the left panel of Fig.\ \ref{deform}.

\item \textbf{Open/ Close}: This update pair constitutes the only possibility to switch between open and closed configurations.
The \textit{Open} move is extecuted as follows:
 \begin{itemize}
 \item Select the time slice of the new head, $\tau_\textnormal{head}$, uniformly from all $3P$ slices.
  \item Select the bead of the new head, $\mathbf{r}_\textnormal{head}$.
\item Select the total number of links to be erased as $m\in[1,\tilde M]$.
\item Select $m$ beads on the next slices from
\begin{eqnarray}
 T_\textnormal{select} = \prod_{i=0}^{m-1} \frac{ \rho_0( \mathbf{r}_i, \mathbf{r}_{i+1}, \epsilon_i) }{ \Sigma_i } \quad ,
\end{eqnarray}
the last one will be the new tail after the update.
\item Delete $m-1$ beads between the new head and tail.
\end{itemize}

The reverse move closes an open configuration. Let $m$ denote the number of missing links between head and tail.
If $m>\tilde M$, the update is rejected.
\begin{itemize}
\item Sample $m-1$ new beads according to Eq.\ (\ref{connect}) with head and tail being the fixed endpoints:
\begin{eqnarray}
 T_\textnormal{sample} = \frac{ \prod_{i=0}^{m-1} \rho_0( \mathbf{r}_i, \mathbf{r}_{i+1}, \epsilon_i )}{ \rho_0(\mathbf{r}_\textnormal{head}, \mathbf{r}_\textnormal{tail}, \epsilon_\textnormal{tot}) } \quad .
\end{eqnarray}
\end{itemize}

The acceptance ratios are computed as
\begin{eqnarray}
 A_\textnormal{Open}(\mathbf{X}\to\mathbf{\tilde X}) &=& \textnormal{min}\left( 1,  \Gamma e^{-\epsilon\Delta\Phi}e^{-\epsilon_\textnormal{tot}\mu} 
\prod_{i=0}^{m-1}\left| \Sigma_i \frac{ \textnormal{det}\rho_i^\textnormal{new}}{ \textnormal{det}\rho_i^\textnormal{old}} \right|  \right) \\
A_\textnormal{Close}(\mathbf{X}\to\mathbf{\tilde X}) &=& \textnormal{min}\left( 1, \frac{ e^{-\epsilon\Delta\Phi}e^{\epsilon_\textnormal{tot}\mu} }{ \Gamma }
\prod_{i=0}^{m-1}\left| \frac{1}{\Sigma_i} \frac{ \textnormal{det}\rho_i^\textnormal{new}}{ \textnormal{det}\rho_i^\textnormal{old}} \right|  \right) \quad , \nonumber 
\end{eqnarray}
with the definition
\begin{eqnarray}
 \Gamma = \frac{3CP\tilde MN}{\rho_0(\mathbf{r}_\textnormal{tail}, \mathbf{r}_\textnormal{head}, \epsilon_\textnormal{tot})} \quad .
\end{eqnarray}
The parameter $\mu$ is another degree of freedom of the algorithm and plays the same role as the chemical potential in the usual WA-PIMC scheme.

\item \textbf{Swap}: The \textit{Swap} move very efficiently generates exchange, i.e., allows for a switch between large off-diagonal or diagonal diffusion matrix elements as it is illustrated in the right panel of Fig.\ \ref{deform}.
Let $m$ denote the number of missing beads between head and tail. 
\begin{itemize}
 \item Choose a target bead on the slice $\tau_\textnormal{tail}$ according to
\begin{eqnarray}
 T_\textnormal{target} = \frac{ \rho_0(\mathbf{r}_\textnormal{head}, \mathbf{r}_t, \epsilon_\textnormal{tot} ) }{ \Sigma_\textnormal{forward} } \quad ,
\end{eqnarray}
with $\Sigma_\textnormal{forward}$ being the normalization. The tail itself cannot be chosen.
\item Choose backwards $m+1$ beads according to
\begin{eqnarray}
 T_\textnormal{select} =  \prod_{i=0}^m \frac{ \rho_0( \mathbf{r}_{i+1}^\textnormal{old}, \mathbf{r}_i^\textnormal{old}, \epsilon_i) }{ \Sigma_i^\textnormal{old} } \quad .
\end{eqnarray}
The head itself cannot be selected on the last slice and the last bead will be the new head after the update.
\item 'Connect' the old head with the target bead by re-sampling the $m$ beads between the slices of head and tail according to
\begin{eqnarray}
 T_\textnormal{sample} = \frac{ \prod_{i=0}^m \rho_0( \mathbf{r}_i^\textnormal{new}, \mathbf{r}_{i+1}^\textnormal{new}, \epsilon_i ) }{ \rho_0( \mathbf{r}_\textnormal{head}, \mathbf{r}_\textnormal{target}, \epsilon_\textnormal{tot}) } \quad .
\end{eqnarray}

\end{itemize}
The update is self-balanced and the acceptance ratio is calculated as
\begin{eqnarray}
 A_\textnormal{Swap}(\mathbf{X}\to\mathbf{\tilde X}) = \textnormal{min}\left( 1, \eta
 \prod_{i=0}^m \left| \frac{ \Sigma_i^\textnormal{old} }{ \Sigma_i^\textnormal{new} } \frac{ \textnormal{det} \rho_i^\textnormal{new} }{ \textnormal{det} \rho_i^\textnormal{old} }
 \right|
 \right) \quad ,
\end{eqnarray}
with the abbreviation
\begin{eqnarray}
 \eta = e^{-\epsilon\Delta\Phi} \frac{ \Sigma_\textnormal{forward} }{ \Sigma_\textnormal{reverse} } \quad ,
\end{eqnarray}
and $\Sigma_\textnormal{reverse}$ being the normalization of the selection of the target bead from the reverse move.

 \item \textbf{Advance/ Recede:} These updates move the head forward (backward) in the imaginary time. However, they are optional and, in principle, not needed for ergodicity.
 The \textit{Advance} move is executed as follows:
 \begin{itemize}
  \item Calculate the number of missing beads between head and tail, $\alpha$. If $\alpha=0$, the update is rejected.
  \item Select the number of new beads to be sampled, $m\in[1,\alpha]$.
  \item Sample the position of the new head from $\rho_0( \mathbf{r}_\textnormal{head}, \mathbf{r}_\textnormal{head}^\textnormal{new}, \epsilon_\textnormal{tot})$.
  \item Sample the $m-1$ beads between old and new head according to Eq.\ (\ref{connect})
  \begin{eqnarray}
   T_\textnormal{sample} = \frac{ \prod_{i=0}^{m-1} \rho_0( \mathbf{r}_i^\textnormal{new}, \mathbf{r}_{i+1}^\textnormal{new}, \epsilon_i) }{ \rho_0(\mathbf{r}_\textnormal{head}, \mathbf{r}_\textnormal{head}^\textnormal{new}, \epsilon_\textnormal{tot} ) } \quad .
  \end{eqnarray}
\end{itemize}
The reverse move is given by \textit{Recede}. Let $\kappa$ denote the total number of beads which can be removed. If $\kappa=0$, the update is rejected.
\begin{itemize}
 \item Select the total number of beads to be removed as $m\in[1,\kappa]$.
 \item Select $m$ beads backwards starting from the old head from 
 \begin{eqnarray}
 T_\textnormal{select} = \prod_{i=0}^{m-1} \frac{ \rho_0(\mathbf{r}_i^\textnormal{new}, \mathbf{r}_{i+1}^\textnormal{new}, \epsilon_i) }{ \Sigma_i^\textnormal{new} } \quad ,
 \end{eqnarray}
 with $\Sigma_i^\textnormal{new}$ being the normalization.
 The last one will be the new head after the update.
 Here ``new'' denotes new with respect to \textit{Advance}, since the coordinates are pre-existing for the \textit{Recede} move. Delete the $m$ beads between the new head and tail.
 \end{itemize}

 This gives the acceptance ratios
\begin{eqnarray}
 A_\textnormal{Advance}(\mathbf{X}\to\mathbf{\tilde X}) &=& \textnormal{min} \left(1, \theta  e^{-\epsilon\Delta\Phi } \prod_{i=0}^{m-1} \left| \frac{1}{\Sigma_i^\textnormal{new}} \frac{ \textnormal{det}\rho_i^\textnormal{new} }{ \textnormal{det}\rho_i^\textnormal{old} } \right|
\right) \\ A_\textnormal{Recede}(\mathbf{X}\to\mathbf{\tilde X}) &=& \textnormal{min}\left(1, \frac{e^{-\epsilon\Delta\Phi}}{\theta}  \prod_{i=0}^{m-1} \left| \Sigma_i^\textnormal{new} \frac{ \textnormal{det}\rho_i^\textnormal{new} }{ \textnormal{det}\rho_i^\textnormal{old} } \right|  \right) \nonumber \quad ,
 \end{eqnarray}
with the definition
\begin{eqnarray}
 \theta = \frac{ \alpha }{ \kappa }  e^{\epsilon_\textnormal{tot}\mu}  \quad .
\end{eqnarray}

\end{enumerate}

 The presented list of Monte Carlo moves constitutes an ergodic set of local updates, which allows for an efficient 
 sampling of both the extended configuration space and a canonical Markov chain.

\end{document}